\documentclass[showpacs,preprintnumbers,amsmath,amssymb]{revtex4}
\usepackage{graphicx}
\usepackage{dcolumn}
\usepackage{epsfig}
\usepackage[dvips]{color}
\usepackage{bm}

\allowdisplaybreaks[4]

\begin{document}
\title{\begin{flushright} {\small KYUSHU-HET 58} \end{flushright}
Oscillation enhanced search for new interaction with neutrinos}

\author{Toshihiko Ota,$^{1}$ \thanks{toshi@higgs.phys.kyushu-u.ac.jp}
        Joe Sato,$^{2}$ \thanks{joe@rc.kyushu-u.ac.jp}
        Nao-aki Yamashita$^{1}$ \thanks{naoaki@higgs.phys.kyushu-u.ac.jp}}

\address{$^1$Department of Physics, Kyushu University,\\ 
         Hakozaki, Higashi-ku, Fukuoka 812-8581, Japan\\
         $^2$Research Center for Higher Education, Kyushu University,\\
         Ropponmatsu, Chuo-ku, Fukuoka 810-8560, Japan}
	 

\begin{abstract}
We discuss the measurement of new physics in long baseline
neutrino oscillation experiments.
Through the neutrino oscillation, 
the probability to detect the new physics effects such as 
flavor violation is 
enhanced by the interference with the weak interaction.
We carefully explain the situations that the interference can take place.
Assuming a neutrino factory and an upgraded conventional beam, 
we estimate the feasibility to observe new physics numerically  
and point out that we can search new interactions 
using some channels, for example
 $\nu_{\mu} \rightarrow \nu_{\mu}$, in these experiments.
We also discuss several models which induce the effective interactions 
interfering with the weak interaction, 
and show that some new physics effects are large enough to 
be observed in the oscillation enhanced way.  
\end{abstract}



\maketitle
\section{Introduction}

Neutrino oscillation induced by light neutrino masses and lepton mixings
gives a plausible explanation for the results of the many neutrino
experiments. 
In the three active neutrino framework, the neutrino mixing is given by,
\begin{equation}
 \nu_{\alpha}  =\sum_{i=1,2,3}U_{\alpha i} \nu_{i}~~(\alpha = e,\mu,\tau),
\end{equation}
where $\alpha$ is the flavor index, $i$ is the mass-eigenstate
index.
The mixing matrix $U$ is a $3\times 3$ unitary matrix, 
called Maki-Nakagawa-Sakata(MNS) matrix \cite{MNS},  defined as 
\begin{equation}
U = \left( \begin{array}{ccc}
      1 & 0 & 0 \\
      0 & c_{23} & s_{23} \\
      0 & - s_{23} & c_{23} \end{array} \right)
\left( \begin{array}{ccc}
      c_{13} & 0 & s_{13} e^{i\delta} \\
      0 & 1 & 0 \\
      -s_{13} e^{-i\delta} & 0 & c_{23} \end{array} \right)
    \left( \begin{array}{ccc}
     c_{12} & s_{12} & 0 \\
      -s_{12} & c_{12} & 0 \\
      0 & 0 & 1 \end{array} \right),
\end{equation}
where $s_{ij}(c_{ij})$ stands for $\sin\theta_{ij}(\cos\theta_{ij})$.
The atmospheric neutrino anomaly \cite{atm} strongly suggests
$\nu_{\mu}\rightarrow\nu_{ \tau}$ oscillation with large mixing,
$4|U_{\mu 3}|^2(1-|U_{\mu 3}|^2) >0.9,$ and a larger mass squared
difference, $\delta m^2_{31}\simeq 2.5\times 10^{-3} {\rm eV}^2$, which
is almost confirmed by the K2K experiment \cite{K2K} and is expected to
be reconfirmed in near future \cite{Minos, OPERA}.  The solar neutrino
deficit \cite{solar} is also explained by the oscillation of $\nu_{e}$
into another neutrino state. It gives several allowed regions for
$|U_{e1}/U_{e2}|$ and the smaller mass squared difference $\delta
m^2_{21}$ among which the region for large mixing MSW \cite{MSW} (LMSW)
solution seems most preferable.  To survey the LMSW region the KamLAND
experiment \cite{KamLand} will start soon. Also the Borexino
experiment \cite{Borexino} will strongly constrain the parameter
region.\footnote{LSND experiment \cite{LSND} also shows a positive signal
for neutrino oscillation which will be tested soon.\cite{MiniBooNE}
However we ignore here.}

However, we do not have sufficient information to determine the all
mixing angles and the sign of the mass squared differences.  $|U_{e3}|$
is constrained strongly by CHOOZ \cite{Chooz} 
and Palo Verde \cite{PaloVerde} experiments;
$|U_{e3}|<0.16$, but it has not determined yet. 
We also have no information about the sign of $\delta
m^2_{31}$ \footnote{Investigation about supernova may already have given
the information about its sign.\cite{supernova}}, and the CP phase $\delta$.
Therefore there are many proposals for neutrino oscillation
experiments
to determine them. The future neutrino-oscillation
experiments based on an accelerator, on both a conventional beam
\cite{conventional} and a muon
storage ring \cite{nuFact}, 
are expected to give high precision tests of oscillation.
Indeed according to these studies, we will determine the mixing angles
and mass squared differences very precisely. The parameters relevant with
atmospheric neutrino anomaly will be determined with error of a few \%.
In these experiments we can explore very small value ($\sim 0.01$)
of $U_{e3}$ and we have a chance to observe the CP violation 
in the lepton sector.

Till today main concern on future neutrino oscillation experiments is
how precisely we can determine the oscillation parameters. Is it all we
can do in such experiments ? 
The answer is NO.\cite{Grossman, NewPhysMatter}
In addition to the
neutrino oscillation parameters we can probe the new lepton-flavor
violating physics.  It affects for example muon decay, matter effect and
so on. Furthermore, as we will see in this paper, the effect of these
exotic couplings is enhanced in oscillation phenomena. Even if the
coupling is rather small, such new physics modifies the oscillation
pattern distinguishably and hence it will be detectable.

In this paper we investigate the possibility to measure the effect of
new physics in future neutrino oscillation experiments. We first study
how the new physics contribute to the neutrino oscillation phenomena in
section II. 
There we will see the reason why its effect is enhanced in
the oscillation physics. In section III, we embody the formalism and adopt
it for a neutrino factory and an upgraded conventional beam.
We present numerical results on the feasibility to search new
interactions in section IV.  
We argue that the energy dependence of event rate at
relatively high energy region is a key issue for a observation. 
We will discuss the relation between lepton flavor violating 
interaction and some models in section
V.  Finally, a summary and discussion are given in section VI.

\section{BASIC IDEA}

In this section we reexamine what is measured in oscillation experiments
and see how new physics gives contribution to the oscillation
phenomena.  Though we discuss only a neutrino factory to make the argument
concrete, same discussion is followed in oscillation experiments with
a conventional beam.

\begin{figure}[h]
\center
\epsfig{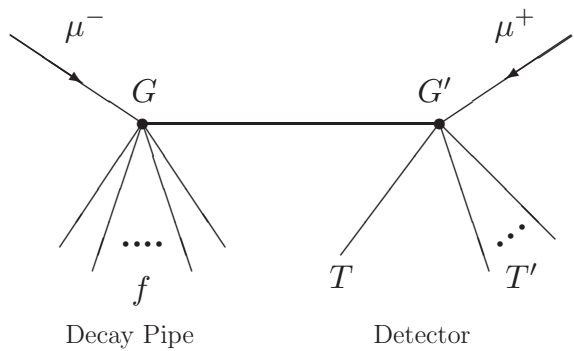}
\caption{What we really see in a neutrino factory.}
\label{fig:weknow}
\end{figure}

First we remind ourselves what is really measured in a  neutrino
factory. All we know is that the muons, say, with negative charge
decay at
an accumulate ring and wrong sign muons are observed in a detector located at a
length $L$ away just after the time $L/c$, where $c$ is the light speed.
This is depicted schematically in Fig.\ref{fig:weknow}.

\begin{figure}[h]
\center
\epsfig{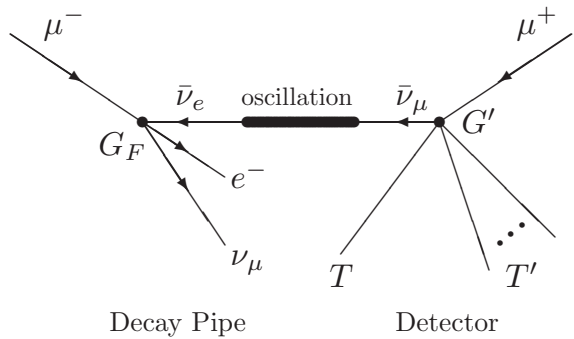}
\caption{Standard interpretation of a wrong sign event.}
\label{fig:weak}
\end{figure}

Since we know that there is the weak interaction process, we interpret
such a wrong sign event as the evidence of the neutrino oscillation,
$\bar\nu_e\rightarrow\bar\nu_\mu$, which is graphically represented in
Fig.\ref{fig:weak}.

\begin{figure}[h]
\center
\epsfig{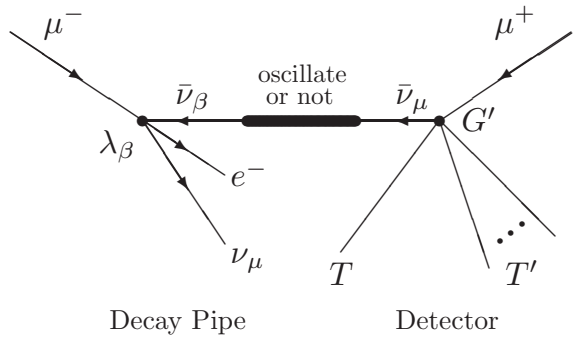}
\caption{Diagram which gives same signal as that given by Fig.\ref{fig:weak}.}
\label{fig:same}
\end{figure}

Now if there is a flavor-changing exotic interaction, {\it e.g.},
\begin{equation}
 \lambda (\bar e \gamma_\mu \mu) (\bar{\nu_\mu} \gamma^\mu \nu_\alpha),~~
\alpha \ne e,
\label{eq:newPhysicsAtCreation}
\end{equation}
then we will have the same signal of a wrong sign muon, whose diagram is shown
in Fig.\ref{fig:same}, just like that caused by the weak interaction 
and the neutrino oscillation.
We cannot distinguish these two kinds of contribution. The quantum mechanics
tells us that in this case, to get a transition rate,  we first sum up
 these amplitudes and then square the summation. 
Therefore there is an interference phenomenon between several amplitudes 
in this process.
Through this interference we get an enhancement of the effect of
new physics, that is, we can make oscillation-enhanced search for new
interactions with neutrinos.

More formally we illustrate this situation as follows.  Denote the
transition amplitude between the initial state $\mathcal A + \mathcal T$ and
the final state $\mathcal C + \mathcal U$ through an intermediate state
$\mathcal B$ as $\Phi({\mathcal A,\mathcal T};{\mathcal B};{\mathcal
C,\mathcal U})$, where a final state $\mathcal U$ is one of the possible
states which are unobserved like $\mathcal B$.  The transition 
probability between $\mathcal A + \mathcal T$ and $\mathcal C + \mathcal
U$, which is the absolute square of the sum of transition amplitudes
with the {\it same} final state, is
\[
 P({\mathcal A + \mathcal T} \rightarrow {\mathcal C + \mathcal U}) = 
\left| \sum_{\mathcal B}
\Phi({\mathcal A, \mathcal T};{\mathcal B};{\mathcal C, \mathcal U}) 
\right|^{2}.
\]
The transition probability from $\mathcal A$ to $\mathcal C $ is given
by summing up all unobserved initial states $\mathcal T$ and final states
$\mathcal U$ as
\begin{eqnarray}
P({\mathcal A} \rightarrow {\mathcal C}) = 
\sum_{\mathcal T,\mathcal U} P({\mathcal A+\mathcal T} \rightarrow {\mathcal
C+\mathcal U}).
\label{eq:formalProbability}
\end{eqnarray}
Furthermore suppose that the amplitude $\Phi({\mathcal A,\mathcal
T_{0}};{\mathcal B_{0}};{\mathcal C,\mathcal U_{0}})$ is dominant
over the amplitudes with intermediate states $\mathcal B ( \neq
\mathcal B_{0})$ and/or initial states $\mathcal T (\neq \mathcal T_{0})$
and/or final states $\mathcal U (\neq \mathcal U_{0})$.  Then the
dominant contribution to the probability 
$P({\mathcal A} \rightarrow {\mathcal C})$\footnote{If the contribution
from $\mathcal T_{0}$ or $\mathcal U_{0}$ is not dominant in
eq.\eqref{eq:formalProbability}, we have to sum up appropriately.} 
is given by $P({\mathcal A} + {\mathcal T}_{0} \rightarrow {\mathcal C}+\mathcal U_{0})$
and deformed as
\begin{eqnarray}
&&P({\mathcal A} \rightarrow {\mathcal C}) \simeq  
 P({\mathcal A + \mathcal T_{0}} \rightarrow \mathcal C +\mathcal U_{0})
\label{eq:dominantContribution} \\
&&~~~~~~~~~~~~~=
 \left| \Phi({\mathcal A, \mathcal T_{0}};{\mathcal B_{0}};\mathcal
C,\mathcal U_{0}) \right|^{2}
\label{eq:leadingTerm}\\
&&~~~~~~~~~~~~~~~~~+ 2 {\rm Re} \left[ \Phi({\mathcal A,\mathcal T_{0}};
{\mathcal B_{0}};\mathcal C,\mathcal U_{0})^{*} 
\sum_{\mathcal B \neq
\mathcal B_{0}} \Phi({\mathcal A,\mathcal T_{0}};
{\mathcal B};\mathcal C,\mathcal
U_{0}) \right]
\label{eq:interferenceTerm}\\
&&~~~~~~~~~~~~~~~~~+ \left|\sum_{{\mathcal B}\neq {\mathcal B}_{0}}
\Phi({\mathcal A,\mathcal T_{0}};{\mathcal B};
\mathcal C,\mathcal U_{0}) \right|^{2}.
\end{eqnarray}
Here the second term in eq.\eqref{eq:interferenceTerm}
is given by the interference among the leading
amplitude $\Phi({\mathcal A,\mathcal T_{0}}; {\mathcal B_{0}};\mathcal
C,\mathcal U_{0})$ and the sub leading amplitudes $\Phi({\mathcal
A,\mathcal T_{0}}; {\mathcal B}(\neq\mathcal B_{0});\mathcal C,\mathcal
U_{0})$.  Note that the interference arises among the processes which
have quite the same initial and final states, even if some of them are
{\it unobserved}.  ``Same state'' means a state with not only the same
particle species but also all other same physical quantities, such as
same energy and same helicity.  This is very important in the
calculation of the transition rate.

The leading amplitude is expected to arise from well known physics.  
On the other hand the amplitudes $\Phi({\mathcal A,\mathcal T_{0}};
{\mathcal B}(\neq \mathcal B_{0});\mathcal C,\mathcal U_{0})$ is
relatively small and may contain the contribution from new physics.
However, even though $|\Phi({\mathcal A,\mathcal T_{0}};{\mathcal
B}(\neq \mathcal B_{0});\mathcal C,\mathcal U_{0})|^{2}$ is very small,
it would be possible for the interference term to be large. Thus even if
the effect of new physics is not detectable due to systematic errors in
a direct measurement, we may see the oscillation-enhanced effect of new
physics.

In neutrino oscillation experiments, the state $\mathcal A$ consists
of the state of the parent particle of neutrinos
and $\mathcal T$ consists from the state of the target particle that
interacts with a neutrino in a detector.  The final state $\mathcal C$
corresponds to the state of the particle which can induce an identifiable
event in a detector as a result of the interaction with the target
particle ($\mathcal T$). Other final states $\mathcal U$ denote the
states of all unobserved particles, which appear at both the
neutrino-creation process and the detection process.  
The intermediate state $\mathcal B$ is interpreted as a neutrino
state. 
The leading amplitude is given by the weak interaction 
and the neutrino oscillation, 
so the final state $\mathcal U_{0}$ is produced through the
weak interaction, 
and the intermediate state $\mathcal B_{0}$ is a neutrino
state which ``oscillates'' from a certain weak interaction eigenstate to
same/another one.  
Even if there is no oscillation,
different intermediate states $\mathcal B(\neq \mathcal B_{0})$ 
induced by exotic interactions 
give non-vanishing amplitudes for the total ``oscillation'' process.
The new physics
would violate the flavor conservation and/or the standard chiral
property.  
Amplitudes which have all the same initial states $\mathcal
T_{0}$ and final states $\mathcal U_{0}$ interfere with each other.  
It means that the leading amplitude from the weak interaction interferes
with  an exotic amplitude.  
In a neutrino-oscillation experiment
 such interference would give sub-leading contribution to the
total probability.  
A large difference from the standard oscillation
will be realized.
\begin{figure}[h]
\center
\epsfig{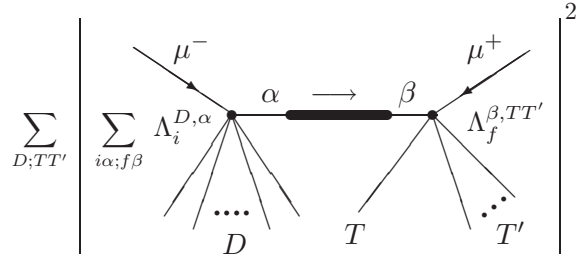}
\caption{Transition rate for ``$\bar\nu_e\rightarrow\bar\nu_\mu$''.}
\label{fig:true}
\end{figure}
\begin{figure}[h]
\center
\epsfig{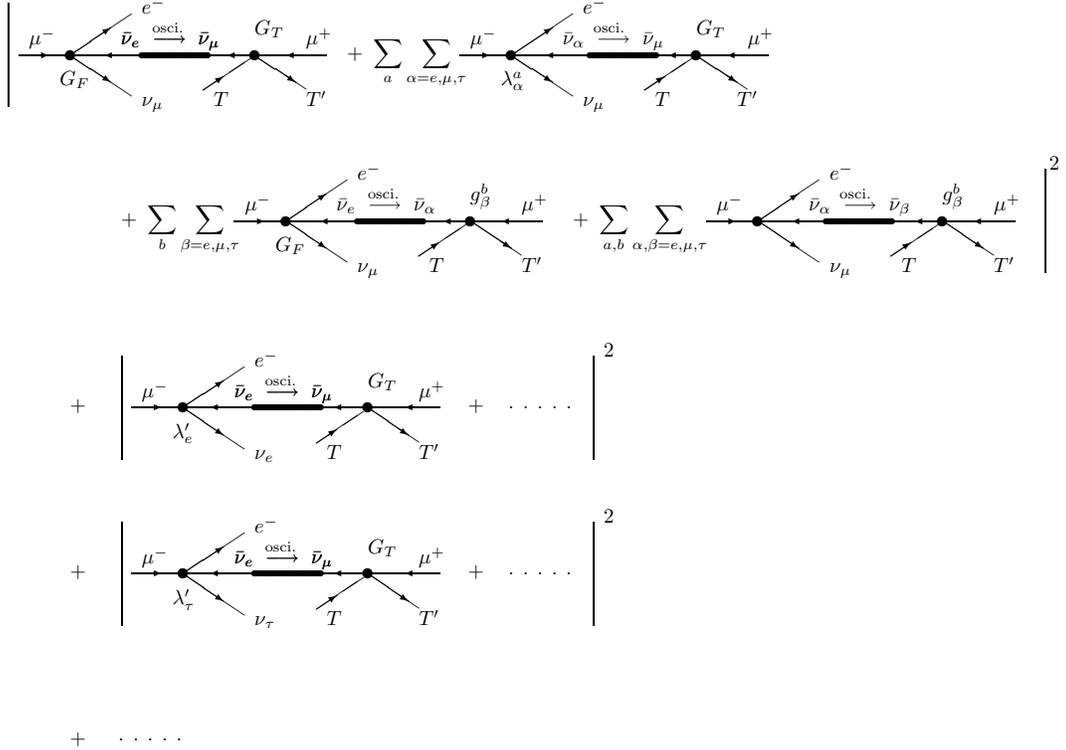}
\caption{
The transition probability of a wrong-sign muon production process
through the neutrino ``oscillation''.
We denote $T$ and $T'$ as the initial and the final state of 
the target particle of the neutrino in a detector.
Large dots describe effective interactions.  
Effective interactions characterized by $G_{F}$ and
$G_{T}$ are induced by the weak interaction. 
We assume that the other effective couplings, 
like $\lambda_{e}^{a}$, are caused by new physics and their superscripts 
identify the types of effective interactions. 
The bold line in each diagram stands for the neutrino oscillation including 
all matter effects induced by the weak interaction and 
exotic interactions. 
Here, appropriate integrations of momenta and summations of helicity
 states for unobserved particles are omitted. 
}
\label{fig:expansion}
\end{figure}
{
\begin{figure}[h]
\center{
\epsfig{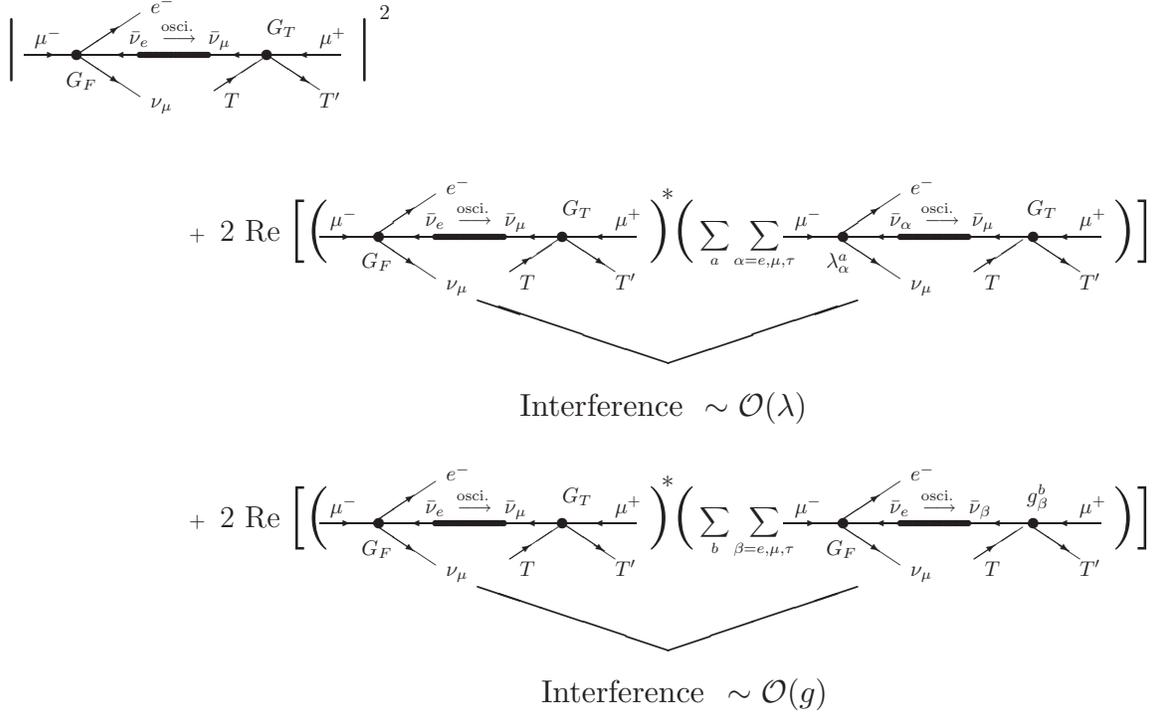}
}
\caption{
The transition probability of a wrong-sign muon appearance process up to 
sub-leading contribution.
}
\label{fig:interference}
\end{figure}
}

To put the situation concretely, we consider a wrong-sign muon
production process through the neutrino oscillation
$\bar\nu_e\rightarrow \bar\nu_\mu$.  
In this case the initial state 
$\mathcal A$ is $ \mu^{-} $ and the observed final state $\mathcal C$
is $\mu^{+}$.  
The above consideration tells that we have to calculate
the transition rate depicted in Fig.\ref{fig:true}. 
Namely eq.\eqref{eq:formalProbability} corresponds to Fig.\ref{fig:true}:
The correspondences for the other symbols are 
$\mathcal T \leftrightarrow T$ and $\mathcal U
\leftrightarrow D + T'$, respectively.  
In this process, the new physics
effects contribute to three parts; the neutrino-creation process as
mentioned above in eq.\eqref{eq:newPhysicsAtCreation}, the detection process,
{\it e.g.},
\begin{equation}
 g (\bar{\nu_e} \gamma_\mu \mu) (\bar{d} \gamma^\mu u),
\label{eq:detector}
\end{equation}
and the matter effect described as
\begin{equation}
 H = 2\sqrt{2}G_F n_e\left( \begin{array}{ccc}
  \epsilon^{m}_{ee} & \epsilon^{m}_{e\mu} & \epsilon^{m}_{e\tau}\\
 \epsilon^{m*}_{e\mu} & \epsilon^{m}_{\mu\mu}& \epsilon^{m}_{\mu\tau}\\
 \epsilon^{m*}_{e\tau} & \epsilon^{m*}_{\mu\tau}& \epsilon^{m}_{\tau\tau}\\
 \end{array} \right),
\label{eq:flavor-changing-matter-effect}
\end{equation}
where $n_{e}$ is the electron number density in matter.\footnote{
Eventually $\epsilon_{ee}^m$ makes a shift on an
observed $n_e$ experimentally.} 
For anti-neutrinos such an exotic matter
effect is given by $-H^*$.  The interaction Hamiltonian of the exotic
matter effect is induced in the similar way to the standard matter
effect.

By expanding Fig.\ref{fig:true}, the transition probability of this
process is illustrated in Fig.{\ref{fig:expansion}}.  In
Fig.{\ref{fig:expansion}} the first diagram on the first line describes
the contribution from the weak interaction and the oscillation process,
where the unobserved state $\mathcal U_{0}$ corresponds to $\{ e^{-}$,
$\nu_{\mu}$ ,$T'\}$.  It gives the leading amplitude.  
The other diagrams on the first line in Fig.\ref{fig:expansion} 
represent the effects of new physics, in which the states of
neutrinos before and/or after the oscillation differ from the weak
interaction case in flavor and/or chiral property.  
Since diagrams in the first line have
completely the same final states $\mathcal C$ and $\mathcal U_{0}$, they
interfere with each other.  
The diagrams on the second and the third line in Fig.\ref{fig:expansion}
have different final states form those of  the first line,
so that there is no interference with the weak interaction   
and it is expected that their contributions to the transition rate is
negligibly small.

Thus eq.\eqref{eq:dominantContribution} is given by the diagram of the
first line in Fig.\ref{fig:expansion}. 
Figure \ref{fig:interference} shows 
the eq.\eqref{eq:leadingTerm} and eq.\eqref{eq:interferenceTerm}.
The leading probability is proportional to $G_{F}^{2}G_{T}^{2}$ since it is
known that exotic effective interactions are so small according to the
direct searches, 
that $|\lambda|^2\ll|G_F|^2$ and $|g|^2\ll|G_T|^2$. 
However, the effect of new physics gives ${\mathcal
O}(\lambda)$ and ${\mathcal O}(g)$ contribution to the signal of wrong
sign muon due to the interference with the amplitude of the weak
interaction.
  This fact makes the search for new physics possible 
in an oscillation-enhanced way.  
It gives relatively large effect. 
On the contrary, even if the
stored muon are very high intense, 
a direct detection of an exotic decay process 
will be very difficult due to systematics 
since the probabilities of such processes are proportional 
to the squares of the effective coupling constants 
and hence is expected too small to be detected.

\section{Formalism and parameterization of an Interference effect with 
new physics in neutrino oscillation}

In this section we consider the neutrino oscillation in the presence of new
physics more concretely to get the analytic expression for the
transition rate. Hereafter we consider only diagrams which interfere
with that of the known physics, {\it i.e.}, the weak interaction.

First we note that the amplitude for ``neutrino oscillation'' can be
divided into three pieces: 
(1) Amplitude relevant with decay of a 
parent particle denoted as  $A^C_\alpha$,
here $C$ describes the type of interactions. 
For $\mu$ decay, as we will see in eq.\eqref{v-av-a} and
\eqref{v-av+a}, there are two types of interactions, $C=L,R$ 
while for $\pi$ decay we do not need this label.  
$\alpha$ distinguishes the
particle species which easily propagates in the matter
 and make an interaction at a detector. 
(2) Amplitude representing a transition of these propagating
particles, which are usually neutrinos, 
from one species $\alpha$ to another/same $\beta$
, denoted as  $T_{\alpha\beta}$. 
(3) Amplitude responsible for producing a charged lepton $\l$ 
from a propagated particle $\beta$ at a
detector, stood for by $D^I_{\beta l}$. Here $I$ denotes an interaction type.
Using these notations we get the probability to observe a charged lepton
$l^\pm$ at a detector as
\begin{eqnarray}
P_{\mu^{-}\rightarrow l^{+}(l^{-})} &=&
 \left|\sum_{\alpha\beta C I}A^C_\alpha
 T_{\alpha\beta} D_{\beta l^\pm}^I \right|^2     \nonumber \\
&=&\sum_{\alpha\beta C I}\sum_{\alpha '\beta ' C' I'}
A^C_\alpha T_{\alpha\beta} D_{\beta l^\pm}^I
A^{C'*}_{\alpha '} T_{\alpha '\beta '}^* D_{\beta 'l^{\pm}}^{I'*}.
\end{eqnarray}
Therefore we can consider the effect of new physics separately for
decay, propagation and detection processes.

First we consider the decay process of parent particles.
Since all final states must be the same, for a neutrino factory, the
exotic decays of muons which are $\mu^{-}\rightarrow
e^{-}\nu_{\alpha}\bar{\nu}_{e}$ and $\mu^{-}\rightarrow
e^{-}\nu_{\mu}\bar{\nu}_{\beta}$ can be amplified by the
interference. 
Though in the presence of Majorana mass terms
neutrinos and anti neutrinos can mix with each other, this effect is
strongly suppressed by $m_{\nu}/E_{\nu}$. 
Therefore we do not have to
consider decays into neutrino with opposite chirality such as
$\mu^{-}\rightarrow e^{-}\bar\nu_{\mu}\bar{\nu}_{\beta}$  The former is
relevant with $\mu^-\rightarrow l^-$ and the latter is relevant with
$\mu^-\rightarrow l^+$. In other words, we can approximate neutrinos to
be massless except for the propagation process.   
This fact and Lorentz invariance allow only two kinds of
new interactions  in this process.  
For a wrong-sign mode, the allowed two
interactions are the $(V-A)(V-A)$ type,
\begin{equation}
2\sqrt{2}\lambda_{\alpha}(\bar{\nu_{\mu}}\gamma^{\rho}P_{L}\mu )
(\bar{e} \gamma_{\rho}P_{L}\nu_{\alpha}),
~~\alpha = \mu,\tau,
\label{v-av-a}
\end{equation}
which has the same chiral property as the weak interaction but violates
the flavor conservation,  and the $(V-A)(V+A)$ type,
\begin{equation}
2\sqrt{2}\lambda'_{\alpha}
(\bar{\nu_{\mu}}\gamma^{\rho}\nu_{\alpha})(\bar{e}\gamma_{\rho} P_{R}\mu ),
~~\alpha = e,\mu,\tau.\label{v-av+a}
\end{equation}
The latter has different chiral property from the former, so that it
gives different energy dependence to the transition rate. 
These exotic interactions interfere with the leading amplitude 
and contribute as next leading effects.  
Note that generally $\lambda$ and $\lambda'$ are
complex numbers.\cite{Grossman}

In the case of the $(V-A)(V-A)$ type exotic interaction, 
we can introduce the interference effect by treating the 
initial state of oscillating neutrino as
 the superposition of all flavor eigenstates.
On the $\mu^{-}\rightarrow \mu^{+}$ process, we can take 
initial neutrino $\bar{\nu}$ as
\begin{eqnarray}
\bar{\nu} = \bar{\nu}_{e} + \epsilon_{\mu} \bar{\nu}_{\mu} 
+ \epsilon_{\tau}\bar{\nu}_{\tau},
\end{eqnarray}
where $\epsilon_{\alpha} = \lambda_{\alpha} / G_{F}$.
This simple treatment is allowed only for the $(V-A)(V-A)$ type interaction 
because of the same interaction form as the weak interaction except for 
difference of the coupling constant and the flavor of antineutrino.
In this case  we can generalize the initial neutrino for any flavor, 
using Y. Grossman's source state notation \cite{Grossman},
as, \footnote{$U^s$ is not necessarily unitary.}
\begin{gather}
 \nu^{s}_{\beta} = U^{s}_{\beta \alpha} \nu_{\alpha} , 
 \quad 
 \alpha, \beta = e, \mu, \tau, 
 \nonumber \\
 U^{s} \equiv 
  \begin{pmatrix}
   1 & \epsilon^{s}_{e \mu} & \epsilon^{s}_{e \tau}    \\
   \epsilon^{s}_{\mu e} & 1 &\epsilon^{s}_{\mu \tau}   \\
   \epsilon^{s}_{\tau e} & \epsilon^{s}_{\tau \mu} & 1
  \end{pmatrix}. 
\label{eq:def-source-state}
\end{gather}
We can include the total exotic effect into the oscillation
probability as
\begin{align}
 P_{\nu^{s}_{\alpha} \rightarrow \nu_{\beta}} &=
  \left|
   \langle \nu_{\beta} | e^{-i H L}
     U^{s}_{\alpha \gamma} | \nu_{\gamma} \rangle  
  \right|^{2}. 
\label{eq:probability-for-V-A}
\end{align}
This treatment is also valid for the effect on the $\nu_{\mu}$
oscillation.

In the case of the $(V-A)(V+A)$ type exotic interaction, we cannot treat
interference terms simply.  The interference term between the weak
interaction and an exotic interaction eq.\eqref{v-av+a}
denoted as $P^{(1)}_{\mu\rightarrow l}$
gives the rate for the observation of the
wrong sign charged lepton, which is interpreted normally as the
oscillation from $\bar\nu_e\rightarrow\bar\nu_l$ in $\mu^-$ decay, as
follows:
\begin{eqnarray}
P^{(1)}_{{\mu}^{-}\rightarrow l^+}&=&\frac{1+\mathcal{P_\mu}}{2}
\frac{1}{(2\pi)^{2}}\sum_{\rm{spin}}\int \frac{d^{3}p_{e}}{2E_{e}}
\frac{d^{3}p_{\nu_{\mu}}}{2E_{\nu_{\mu}}}
\delta^{4}(p_{\mu}- p_{\bar{\nu}} - p_{e}-p_{\nu_{\mu}}) \nonumber \\
&&~~~~~~~~~\times 
 2 {\rm Re}\left[ A_{e}^{L*} \sum_{\beta I}T_{e\beta}^{*}D^{I*}_{\beta l^{+}}
\sum_{\alpha\beta ' I'}
A_{\alpha}^{R}T_{\alpha\beta '}D_{\beta ' l^{+}}^{I'} \right] \nonumber \\
&=&\frac{1+\mathcal{P_\mu}}{2}
\frac{8 G_{F}}{\pi}m_{e}m_{\mu}
E_\nu
(|{\mathbf p}_{\mu}| -E_{\mu}) 
\sum_{\alpha\beta\beta ' I I'}
 {\rm Re}[ \lambda_{\alpha}' 
T_{e\beta}^{*}D^{I*}_{\beta l^{+}} T_{\alpha\beta '}D_{\beta ' l^{+}}^{I'}],
\label{eq:cross-amp}
\end{eqnarray}
where 
$\mathcal P_\mu$ is the polarization of the initial $\mu^-$, $E_\nu$ is
$\nu$ energy, and ${\mathbf p}_\mu (E_\mu)$ is $\mu$ momentum (energy). 

In the case for the observation of the same sign charged lepton, which
is interpreted as the oscillation from $\nu_\mu\rightarrow\nu_l$ in
$\mu^-$ decay, as follows:
\begin{eqnarray}
P^{(1)}_{{\mu}^{-}\rightarrow l^-}&=&
\frac{1-\mathcal P_\mu}{2}\frac{8 G_{F}}{\pi}m_{e}m_{\mu}
E_\nu
(|{\mathbf p}_{\mu}| -E_{\mu}) 
\sum_{\alpha\beta\beta ' I I'}
 {\rm Re}[ \lambda_{\alpha}' 
T_{\mu\beta}^{*}D^{I*}_{\beta l^{-}} T_{\alpha\beta '}D_{\beta ' l^{-}}^{I'}],
\label{eq:cross-amp-numu-forward}
\end{eqnarray}

For $\pi$ decay the situation is much simpler. In the presence of new physics
there may be a flavor violating decay of $\pi$ such as
$\pi^-\rightarrow\mu^-\nu_\alpha (\alpha=e,\tau)$. This effect changes
the initial $\nu$ state;
\begin{eqnarray}
 \nu_\mu\longrightarrow\nu^s_\mu=
\epsilon_{\mu e}^{s} \nu_e+\nu_\mu+ \epsilon_{\mu\tau}^{s} \nu_\tau.
\end{eqnarray}
In this case we do not have to worry about the type of new physics
which gives a flavor changing $\pi$ decay at a low energy scale.
Due to kinematics, the energy and the helicity of the decaying particles, $\mu$
and $\nu$ are fixed.

Next we consider the propagation process.
Exotic interactions also modify the Hamiltonian for neutrino propagation
as \cite{NewPhysMatter},
\begin{align}
H_{\beta \alpha} 
&= \frac{1}{2 E_{\nu}}
     \left\{
       U_{\beta i}
        \begin{pmatrix}
            0 & & \\
              & \delta m_{21}^{2} & \\
              &                   & \delta m_{31}^{2}
        \end{pmatrix}
       U^{\dagger}_{i \alpha}
              +    
       \begin{pmatrix}
         \bar{a}+ a_{ee}  
                & a_{e \mu}   
                & a_{e \tau}  \\
              {a_{e \mu} }^{*}  
                & a_{\mu \mu}  
                & a_{\mu \tau}  \\
              {a_{e \tau} }^{*} 
                & {a_{\mu \tau} }^{*}  
                & a_{\tau \tau} 
        \end{pmatrix}_{\beta \alpha}
       \right\},
\label{eq:Hamiltonian}
\end{align}
where 
$\bar{a}$ is the ordinary matter effect given by
$2\sqrt{2} G_{F} n_{e} E_{\nu}$, 
$a_{\alpha \beta}$ is the extra matter effect due to new
physics interactions, that is defined by $ a_{\alpha \beta} = 2\sqrt{2}
\epsilon^{m}_{\alpha \beta} G_{F} n_{e} E_{\nu} $.  
Note that to
consider the magnitude of the matter effect, 
the type of the interaction is irrelevant 
since in matter particles are at rest and hence the
dependence on the chirality is averaged out.\cite{BGE}

Finally we make a comment about new physics which affect a detection
process. To consider this process we need the similar treatment to 
that at the decay process, that is, 
we have to separate contribution of new interactions
following the difference of the chirality dependence. 
However to take into account new physics at a detector, 
the parton distribution and a knowledge about hadronization are
necessary.
Though we may wonder whether we can parameterize the effect of new
physics at the detector $g/G_{T}$ as $\epsilon^{d}$ like $\epsilon^{s}$.
It is expected that $\epsilon^{d}$ has a complicated energy dependence
due to the parton distribution for example in a energy region of a
neutrino factory.
Consider the case that there is an elementary process from 
lepton flavor violating new physics including strange quark. 
To parameterize its effect we need both its magnitude and the
distribution function of strange quark in nucleon which will show
the dependence on the neutrino energy (more exactly, the transfered
momentum from neutrino to strange quark).
They are beyond our ability and hence we do not consider them further 
in this paper, though new physics which can affect the decay process
 have  contribution to the detection process too.

\section{numerical analysis}

Let us discuss the feasibility to observe the signal induced by new physics.
We will deal with both an upgraded conventional beam and a neutrino factory
but present a little bit qualitatively different analysis in each experiment.
 
For a neutrino factory, we use following procedures:
First, we assume the magnitude of the effects caused by new interactions.
Next, we calculate the event numbers including the effects of new physics 
$N^{NP}$ and also calculate that based on the standard model $N^{SM}$.
Then, we define the following quantity, so called $\chi^{2}$ function
\footnote{Strictly speaking, this quantity is ``power of test'' to
distinguish two theories, the theory including new physics and the standard
model. 
Furthermore, note that since we are interested in ``goodness of fit'' 
for two theories, we should compare this quantity with $ \chi^{2} $ 
distribution function with one degree of freedom.
More detail on statistics is found in Ref.\cite{KOS}.},
\begin{align}
\chi^{2} 
&\equiv
 \sum_{i}^{\rm bin} \frac{\left| N^{NP}_{i} - N^{SM}_{i} \right|^{2}}
                         {N^{SM}_{i}}
 \nonumber \\
&\equiv N_{\mu} M_{\rm det} X^{2}_{\rm \nu-fact},
\label{eq:chi2-def} 
\end{align}
where $i$ is the energy bin index, $N_{\mu}$ is the muon number, and
$M_{\rm det}$ is the detector mass.
To claim that new physics effects can be observed at 90\% confidence level, 
it is required
\begin{align}
\chi^{2} > \chi^{2}_{90\%},
\end{align}
and this condition is rewritten as
\begin{align}
N_{\mu} M_{\rm det} > \frac{\chi^{2}_{90\%}}{X^{2}_{\rm \nu-fact}}.
\label{eq:observe-conditon}
\end{align}
From the above method, we can obtain the necessary muon number 
and the detector mass to observe the new physics effects 
at 90\% confidence level.

On the other hand, there are some kinds of option on beam configurations 
for an upgraded conventional beam; 
wide band, narrow band, off axis, and so on.
We discuss the event number of {\it no-oscillated neutrino events} 
at the detector $N_{\nu}$ unlike the case for a neutrino factory.
The concrete procedure is almost the same as that for a neutrino factory.
We separate $\chi^{2}$ function defined in first line of 
eq.\eqref{eq:chi2-def} into two parts; 
\begin{align}
 \chi^{2} \equiv N_{\nu} X^{2}_{\rm conv},
\end{align} 
and get the necessary number of no-oscillated neutrino events from
\begin{align}
 N_{\nu} > \frac{\chi^{2}_{90\%}}{X^{2}_{\rm conv}}.
\end{align} 
In the numerical calculation for a conventional beam, 
we consider the wide-band-beam like situation, 
whose flux distribution for energy is constant.

\subsection{$(V-A)(V-A)$ type new interaction}
Here, we deal with the case that there are only $(V-A)(V-A)$ type new
interactions in the lepton sector.
In this case, we need to consider the effect represented in 
eqs.\eqref{eq:def-source-state} and \eqref{eq:Hamiltonian}.
Before making the presentation of the numerical calculations, 
we give the analytic expression for the sensitivities to understand the 
essential features.
As we showed in section III, 
the interference terms between $(V-A)(V-A)$ type interactions and 
the weak interaction have the same dependence on the $\mu$ polarization 
leading term does.
It can not be expected that the sensitivity to such interference term
becomes better by the control of the parent $\mu$ polarization.
We consider here an unpolarized muon beam.
The ``oscillation probability'' is given by
eq.\eqref{eq:probability-for-V-A} in this situation.  
More detailed calculations are presented in the Appendix.

\subsubsection{
$ \nu_{e} \rightarrow \nu_{\mu}$ channel in a neutrino factory}

The analytic expression of probability for $\nu_{e} \rightarrow
\nu_{\mu}$ given in the Appendix A shows that 
the effect due to $\epsilon^{m}_{\mu \tau}$ and 
$\epsilon^{s,m}_{\alpha \alpha}$ are irrelevant since these terms are
proportional to $\sin^{2} 2 \theta_{13} \times \epsilon$
in the high energy region, so it is difficult to observe their effects.
The flavor changing processes between muon and electron, {\it e.g.,} 
$\mu \rightarrow e \gamma$, $\mu \leftrightarrow e$ conversion, 
are strictly constrained from experiments, 
and as we argue in the next section the box diagrams of 
the $\mu$-to-$e$ processes must relate to $\epsilon_{e \mu}^{s}$ 
and $\epsilon_{\mu e}^{s}$.
Therefore, the magnitude of $\epsilon^{s}_{e \mu}$ has very severe bound
and the terms depending on it are also not effective.
Assuming some models, {\it e.g.}, MSSM with right-handed neutrinos, 
it is expected that the magnitude of $\epsilon^{s}_{\alpha \beta}$ and 
$\epsilon^{m}_{\alpha \beta}$ are the same order 
because they are produced by similar diagrams.
This is also discussed in the next section.
We assume naively that the terms depending on 
$\epsilon^{m}_{e \mu}$ are constrained as $\epsilon^{s}_{e \mu}$. 
On the other hand, the $\tau$-to-$e$ processes do not 
give the tight bound to $\epsilon_{e \tau}^{s,m}$.
Hence, we investigate the effect induced by $\epsilon^{s,m}_{e \tau}$
first.

Before surveying the required $N_\mu M_{\rm det}$ for each baseline $L$
and muon energy $E_\mu$, 
we see the behavior of contribution of $\epsilon_{e\tau}^{s,m}$ 
to the ``oscillation probability'' 
to consider the optimum setup for $L$ and $E_\mu$.
In the high energy region such as the matter effect $\bar{a}$ is much greater 
than $\delta m_{31}^{2}$, 
the first order contribution of $\epsilon^{s,m}_{e \tau}$ to the
transition probability,
$\Delta P_{\nu_{e} \rightarrow \nu_{\mu}} \{ \epsilon _{e \tau} \}$,
is 
constructed by four parts that have the different $\epsilon_{e \tau}^{s,m}$ 
dependences:
\begin{widetext}
\begin{subequations}
\label{eq:HighEnergy_nuenumu_ETau}
\begin{align}
\Delta P_{\nu_{e} \rightarrow \nu_{\mu}}\{\epsilon_{e \tau}\}
&=
2 s_{23} s_{2 \times 23} s_{2 \times 13} \nonumber \\
& \quad 
 \times
 \Biggl[
   c_{13}^{2}
    \left( s_{\delta} {\rm Re} [\epsilon_{e \tau}^{s}] 
          - c_{\delta} {\rm Im} [\epsilon_{e \tau}^{s}]
    \right)
    \left(
      \frac{\bar{a}}{4 E_{\nu}} L
    \right)
    \left(
     \frac{\delta m_{31}^{2}}{4 E_{\nu}} L
    \right)^{2}
 \label{eq:HighEnergy_nuenumu_ETau_s_ImCosDelta} \\
& \qquad
 +
   c_{13}^{2} 
      \left(
        c_{\delta} {\rm Re}[\epsilon^{s}_{e \tau}]
        + s_{\delta} {\rm Im}[\epsilon^{s}_{e \tau}]
      \right)
    \left\{
      1- \frac{1}{2} \left( \frac{\bar{a}}{4 E_{\nu}} L \right)^{2}
      - s_{13}^{2} \left( \frac{\bar{a}}{4 E_{\nu}} L \right)
    \left( \frac{\delta m_{31}^{2}}{4 E_{\nu}} L \right)  
    \right\} \left( \frac{\delta m_{31}^{2}}{4 E_{\nu}} L \right)^{2} 
 \label{eq:HighEnergy_nuenumu_ETau_s_ReCosDelta}    \\
& \qquad
 - 
  c_{13}^{2}
  \left(
    s_{\delta} {\rm Re}[\epsilon_{e \tau}^{m}] 
    + c_{\delta} {\rm Im}[\epsilon_{e \tau}^{m}]
  \right)
  \left( \frac{\bar{a}}{4 E_{\nu}} L \right)
  \left(
   \frac{\delta m_{31}^{2}}{4 E_{\nu}} L 
  \right)^{2}
  \label{eq:HighEnergy_nuenumu_ETau_m_ImCosDelta}\\
& \qquad
 -
 \frac{1}{3} s_{13}^{2} 
 \left(
  c_{\delta} {\rm Re} [\epsilon_{e \tau}^{m}] 
  - s_{\delta} {\rm Im} [\epsilon_{e \tau}^{m}]    
 \right) 
 \left\{
    \left( \frac{\bar{a}}{4 E_{\nu}} L \right)
    + 2  \left( \frac{\delta m_{31}^{2}}{4 E_{\nu}} L \right)
  \right\}
  \left( \frac{\bar{a}}{4 E_{\nu}} L \right) 
  \left( \frac{\delta m_{31}^{2}}{4 E_{\nu}} L \right)^{2}
 \Biggr]
\label{eq:HighEnergy_nuenumu_ETau_m_ReCosDelta},
\end{align} 
\end{subequations}
\end{widetext}
where $s_{2 \times ij} \equiv \sin 2 \theta_{ij}$.
Since we can suppose that $E_{\nu}$ is proportional to $E_{\mu}$ 
in a neutrino factory, 
$E_{\mu}$ and $L$ dependence of the sensitivity to each term can be
approximated as
\begin{align}
\chi^{2} \eqref{eq:HighEnergy_nuenumu_ETau_s_ReCosDelta} 
\propto
 \left\{
  1 - \frac{1}{2} \left( \frac{\bar{a}}{4E_{\mu}} L \right)^{2} 
 \right\}^{2} \times E_{\mu}, 
\quad
\chi^{2} ({\rm \ref{eq:HighEnergy_nuenumu_ETau_s_ImCosDelta},
               \ref{eq:HighEnergy_nuenumu_ETau_m_ImCosDelta}}) 
\propto
 \left( \frac{\bar{a}}{4 E_{\mu}} L \right)^{2} \times E_{\mu},
\quad  
\chi^{2} \eqref{eq:HighEnergy_nuenumu_ETau_m_ReCosDelta}
\propto 
  \left( \frac{\bar{a}}{4 E_{\mu}} L \right)^{4} \times E_{\mu}.
\label{eq:HighEnergy_nuenumu_ETau_chi2} 
\end{align}
All of eq.\eqref{eq:HighEnergy_nuenumu_ETau_chi2} are proportional to 
$E_{\mu}$,
so the sensitivities must get better as the energy becomes higher.

Each of eq.\eqref{eq:HighEnergy_nuenumu_ETau_chi2} depends on $L$ in
different way. 
$\chi^{2} \eqref{eq:HighEnergy_nuenumu_ETau_s_ReCosDelta}$ become tiny
for longer baseline length within the region that we are now interested in. 
This fact means that a shorter baseline experiment has an advantage over 
a longer one to observe \eqref{eq:HighEnergy_nuenumu_ETau_s_ReCosDelta}'s 
effect.
By contrast with this, it is found that longer baseline will be better
to search for the effects of 
\eqref{eq:HighEnergy_nuenumu_ETau_s_ImCosDelta}, 
\eqref{eq:HighEnergy_nuenumu_ETau_m_ImCosDelta} and 
\eqref{eq:HighEnergy_nuenumu_ETau_m_ReCosDelta}. 

Each of eq.\eqref{eq:HighEnergy_nuenumu_ETau_chi2} depends on 
the combinations of $\epsilon_{e \tau}^{s,m}$ and the CP phase $\delta$.
What we observe is the combination of them. 
The effects of $\epsilon$'s can be sources of the CP-violation 
effect.\cite{Grossman} 
In the discussion about the observation of the CP phase, 
this fact should be considered.
The analytic expressions also show that the sensitivities are proportional to 
$|\epsilon|^2$.

Now, we show the results of the numerical calculations.
The parameters that we use here are
\begin{gather}
\sin \theta_{12} = \frac{1}{2}, \quad
\sin \theta_{23} = \frac{1}{\sqrt{2}}, \quad
\sin \theta_{13} = 0.1, \nonumber \\
\delta m_{21}^{2} = 5 \times 10^{-5}, \quad
\delta m_{31}^{2} = 3 \times 10^{-3}, \\
\delta = \frac{\pi}{2},\nonumber
\end{gather}
and take $|\epsilon| = 3 \times 10^{-3}$, which is a reference value
for the feasibility to observe the effect 
by using the method of the oscillation enhancement.
Except for $\epsilon_{e \mu}^{s,m}$ and $\epsilon_{\mu e}^{s}$,
the constraints of the processes of charged lepton have not forbidden
this magnitude of $\epsilon$'s.

\begin{figure*}[th]
\unitlength=1cm
\begin{picture}(16,4)
\put(1.75,-0.4){$E_{\mu}$[GeV]}
\put(5.8,-0.4){$E_{\mu}$[GeV]}
\put(9.8,-0.4){$E_{\mu}$[GeV]}
\put(13.85,-0.4){$E_{\mu}$[GeV]}
\put(-0.5,1.3){\rotatebox{90}{$L$[km]}}
\includegraphics[width=16cm]{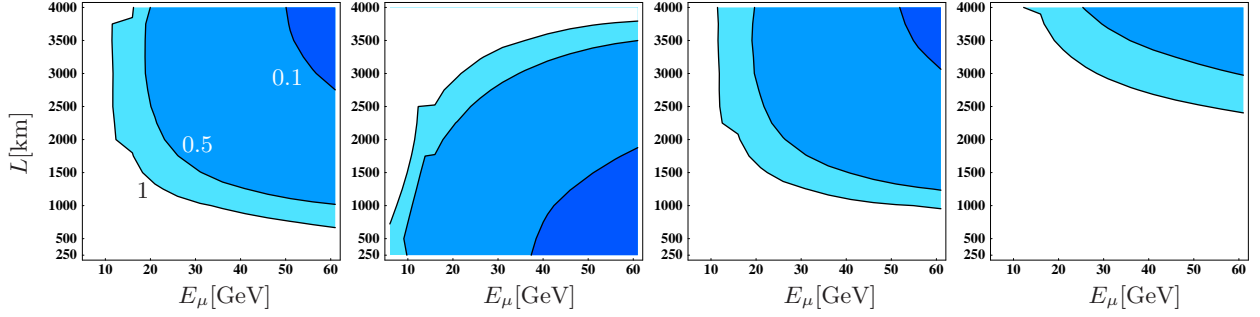}
\put(-14.8,1){1}
\color[rgb]{1,1,1}
\put(-14.2,1.6){0.5}
\put(-13,2.5){0.1}
\color[rgb]{0,0,0}
\end{picture}
\vspace{0.3cm}
\caption{Contour plots of the required $N_{\mu} M_{\rm det}$
to observe the new 
physics effects concerning $\epsilon_{e \tau}^{s,m}$ at 90\% C.L. in 
$\nu_{e} \rightarrow \nu_{\mu}$ channel using a neutrino factory.
From left to right: 
$(\epsilon_{e \tau}^{s}, \epsilon_{e \tau}^{m}) = 
 (3.0 \times 10^{-3}, 0)$, 
$(3.0 \times 10^{-3} i, 0)$, 
$(0, 3.0 \times 10^{-3})$, 
$(0, 3.0 \times 10^{-3} i)$.
The uncertainties of theoretical parameters are not considered in these
plots.
As we point out in the text, when the uncertainties are taken into account,
 the sensitivities are completely lost.
In Fig.\ref{fig:nuenumu_ETau_U0} to Fig.\ref{fig:numunue_EMu},
contours mean 0.01, 0.05, 0.1, 0.5, 1$\times 10^{21} \cdot 100{\rm kt}$.}
\label{fig:nuenumu_ETau_U0}
\end{figure*}

\begin{figure*}[th]
\unitlength=1cm
\begin{picture}(16,8)
\put(1.75,-0.4){$E_{\mu}$[GeV]}
\put(5.8,-0.4){$E_{\mu}$[GeV]}
\put(9.8,-0.4){$E_{\mu}$[GeV]}
\put(13.85,-0.4){$E_{\mu}$[GeV]}
\put(-0.5,1.4){\rotatebox{90}{$L$[km]}}
\put(-0.5,5.4){\rotatebox{90}{$L$[km]}}
\includegraphics[width=16cm]{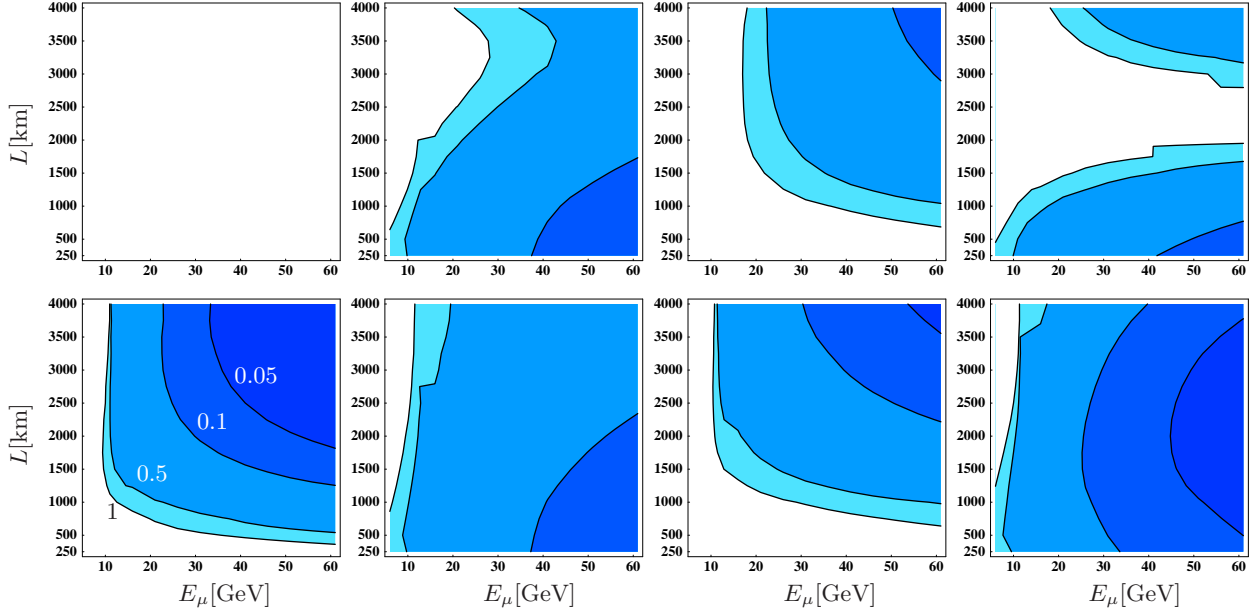}
\put(-15.2,0.7){1}
\color[rgb]{1,1,1}
\put(-14.8,1.2){0.5}
\put(-14,1.9){0.1}
\put(-13.5,2.5){0.05}
\color[rgb]{0,0,0}
\end{picture}
\vspace{0.3cm}
\caption{
Same as Fig.\ref{fig:nuenumu_ETau_U0}, but here
new interactions in both source and matter are taken into account.
From left to right and top to bottom:
$(\epsilon_{e \tau}^{s}, \epsilon_{e \tau}^{m}) = 
 (3.0 \times 10^{-3},   3.0 \times 10^{-3})  $, 
$(3.0 \times 10^{-3} i, 3.0 \times 10^{-3} i)$, 
$(3.0 \times 10^{-3},   3.0 \times 10^{-3} i)$, 
$(3.0 \times 10^{-3} i, 3.0 \times 10^{-3} ) $;
$(3.0 \times 10^{-3},   -3.0 \times 10^{-3})  $, 
$(3.0 \times 10^{-3} i, -3.0 \times 10^{-3} i)$, 
$(3.0 \times 10^{-3},   -3.0 \times 10^{-3} i)$, 
$(3.0 \times 10^{-3} i, -3.0 \times 10^{-3} ) $. 
These plots can be understood by addition or subtraction of the plots 
in Fig.\ref{fig:nuenumu_ETau_U0}.}
\label{fig:nuenumu_ETau_source+matter_U0}
\end{figure*}
Fig.\ref{fig:nuenumu_ETau_U0} shows the required $N_\mu M_{\rm
det}$ in the case where we do not take into account the uncertainties 
of the mixing parameters.
We can check whether the approximated equations, 
eq.\eqref{eq:HighEnergy_nuenumu_ETau_chi2},  are correct 
from the behavior of the plots.  
As eq.\eqref{eq:HighEnergy_nuenumu_ETau} implies,
contribution from the new interaction depends on the combinations of
$\epsilon$ and $\delta$.  
We take $\delta = \pi / 2$, 
so we can extract each term of eq.\eqref{eq:HighEnergy_nuenumu_ETau} 
by taking $\epsilon_{e \tau}^{s,m}$ pure real and imaginary. 
Therefore, the plots of Fig.\ref{fig:nuenumu_ETau_U0} from left to right 
correspond to the required data size to observe each term in 
eq.\eqref{eq:HighEnergy_nuenumu_ETau_s_ImCosDelta} to
eq.\eqref{eq:HighEnergy_nuenumu_ETau_m_ReCosDelta} respectively.  
The behavior of these plots is consistent with the expectations 
from the analytic
expressions in  eq.\eqref{eq:HighEnergy_nuenumu_ETau_chi2}. 
To consider realistic situations, the new physics effects in both 
source and matter must be taken into account simultaneously.  
We present Fig.\ref{fig:nuenumu_ETau_source+matter_U0} as the plots 
in the cases where the same magnitude of new effects 
in both source and matter are
introduced. They show that the total effects are given by the simple
summation of each effect.

\begin{figure*}[th]
\unitlength=1cm
\begin{picture}(16,4)
\put(1.75,-0.4){$E_{\mu}$[GeV]}
\put(5.8,-0.4){$E_{\mu}$[GeV]}
\put(9.8,-0.4){$E_{\mu}$[GeV]}
\put(13.85,-0.4){$E_{\mu}$[GeV]}
\put(-0.5,1.3){\rotatebox{90}{$L$[km]}}
\includegraphics[width=16cm]{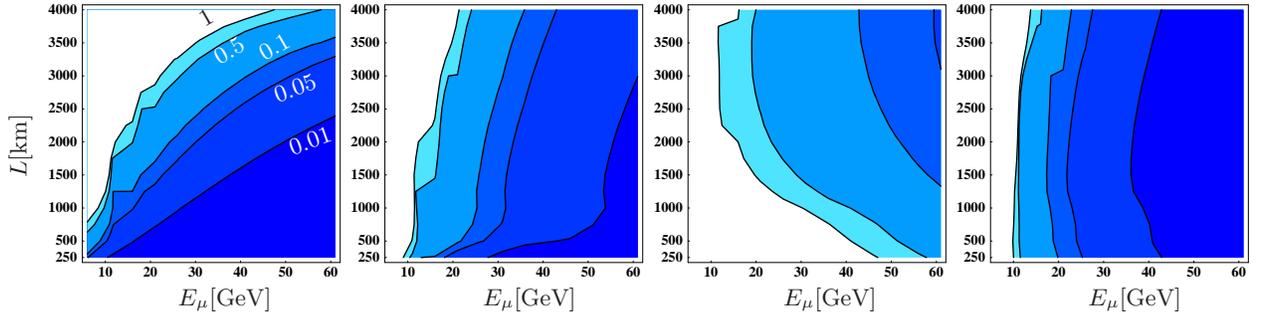}
\put(-14,3.3){\rotatebox{30}{1}}
\color[rgb]{1,1,1}
\put(-13.8,2.8){\rotatebox{30}{0.5}}
\put(-13.2,2.85){\rotatebox{30}{0.1}}
\put(-13,2.3){\rotatebox{20}{0.05}}
\put(-12.8,1.6){\rotatebox{20}{0.01}}
\color[rgb]{0,0,0}
\end{picture}
\vspace{0.3cm}
\caption{Contour plots of the required $N_{\mu} M_{\rm det}$ to observe 
the new physics effects concerning $\epsilon_{e \mu}^{s,m}$ 
at 90\% C.L.
in $\nu_{e} \rightarrow \nu_{\mu}$ channel 
when there is no uncertainty for theoretical parameters.
From left to right: 
$(\epsilon_{e \mu}^{s}, \epsilon_{e \mu}^{m}) = 
 (3.0 \times 10^{-3},   0)                   $, 
$(3.0 \times 10^{-3} i, 0)                   $, 
$(0,                    3.0 \times 10^{-3})  $, 
$(0,                    3.0 \times 10^{-3} i)$. 
Each plot corresponds to the sensitivities to
eq.\eqref{eq:HighEnergy_nuenumu_EMu_s_ImCosDelta}%
$\sim$%
eq.\eqref{eq:HighEnergy_nuenumu_EMu_m_ReCosDelta} respectively.}
\label{fig:nuenumu_EMu_U0}
\end{figure*}

\begin{figure*}[th]
\unitlength=1cm
\begin{picture}(16,4)
\put(1.75,-0.4){$E_{\mu}$[GeV]}
\put(5.8,-0.4){$E_{\mu}$[GeV]}
\put(9.8,-0.4){$E_{\mu}$[GeV]}
\put(13.85,-0.4){$E_{\mu}$[GeV]}
\put(-0.5,1.3){\rotatebox{90}{$L$[km]}}
\includegraphics[width=16cm]{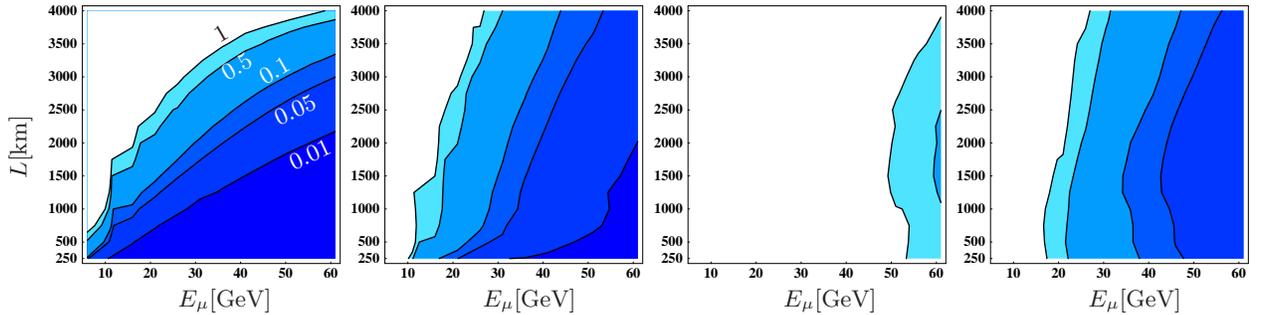}
\put(-13.8,3.1){\rotatebox{30}{1}}
\color[rgb]{1,1,1}
\put(-13.7,2.6){\rotatebox{30}{0.5}}
\put(-13.2,2.55){\rotatebox{30}{0.1}}
\put(-13,2){\rotatebox{25}{0.05}}
\put(-12.8,1.4){\rotatebox{25}{0.01}}
\color[rgb]{0,0,0}
\end{picture}
\vspace{0.3cm}
\caption{Same as Fig.\ref{fig:nuenumu_EMu_U0}, but here each parameter has
 10\% uncertainty.}
\label{fig:nuenumu_EMu_U10}
\end{figure*}

In realistic situations the uncertainties of the theoretical
parameters have to be taken into account.\footnote{
We understand that systematic errors should be also
taken into account. However in this paper we refer to the errors 
originated from statistics and we treat the errors in the
theoretical parameters.}
Once the uncertainties are introduced, it can be expected that 
the sensitivities shown in Figs.\ref{fig:nuenumu_ETau_U0} and 
\ref{fig:nuenumu_ETau_source+matter_U0} will be spoiled completely.
The $\epsilon_{e \tau}^{s,m}$ effect can be absorbed easily into 
the main (unperturbed) part of oscillation 
by adjusting the theoretical parameters 
since the effects have the same energy dependence as the main part has.
Indeed, taking into account these uncertainties, the sensitivities to
$\epsilon_{e \tau}^{s,m}$ are completely washed out.
Therefore, we have to look for the terms whose energy 
dependence differ from that of the main oscillation term 
in the high energy region.
In $\nu_{e} \rightarrow \nu_{\mu}$ channel, 
the effects caused by $\epsilon_{e \mu}^{s,m}$ have such energy dependence.
It can be represented analytically in the high energy region as
\begin{widetext}
\begin{subequations}
\label{eq:HighEnergy_nuenumu_EMu}
\begin{align}
\Delta P_{\nu_{e}^{s} \rightarrow \nu_{\mu}}\{\epsilon_{e \mu}\}
&=
2
s_{23} s_{2 \times 13} \nonumber \\
& \quad
\times
 \Biggl[
  \left(
   s_{\delta} {\rm Re}[\epsilon^{s}_{e \mu}]
   - c_{\delta} {\rm Im}[\epsilon^{s}_{e \mu}]
  \right) \nonumber \\
& \qquad \quad 
  \times
 \Biggl\{
  1 - \frac{2}{3} \left( \frac{\bar{a}}{4 E_{\nu}} L \right)^{2}
  +
  \frac{2}{3}
   \left(
    2 c_{2 \times 13} - 3 c_{23}^{2} c_{13}^{2} 
   \right)
  \left( \frac{\bar{a}}{4 E_{\nu}} L \right)
  \left( \frac{\delta m_{31}^{2}}{4 E_{\nu}} L \right)
  \Biggr\}
  \left( \frac{\delta m_{31}^{2}}{4 E_{\nu}} L \right)
\label{eq:HighEnergy_nuenumu_EMu_s_ImCosDelta} \\
& \qquad
- 
 \left(
        c_{\delta} {\rm Re}[\epsilon^{s}_{e \mu}]
        + s_{\delta} {\rm Im}[\epsilon^{s}_{e \mu}]
      \right) \nonumber \\
& \qquad \quad
  \times
  \Biggl[
   \left\{ 
     1 - \frac{1}{3} \left( \frac{\bar{a}}{4 E_{\nu}} L \right)^{2} 
   \right\}
   \left( \frac{\bar{a}}{4 E_{\nu}} L \right) \nonumber \\
& \qquad \qquad \qquad
  -
   \left\{
    1 - 2 s_{23}^{2} c_{13}^{2}
    -
     \left(
      1 - c_{13}^{2} \left(2 - \frac{4}{3} c_{23}^{2} \right)
     \right)
     \left(
      \frac{\bar{a}}{4 E_{\nu}} L
     \right)^{2}
   \right\}
   \left( \frac{\delta m_{31}^{2}}{4 E_{\nu}} L \right)
  \Biggr]
  \left( \frac{\delta m_{31}^{2}}{4 E_{\nu}} L \right) 
\label{eq:HighEnergy_nuenumu_EMu_s_ReCosDelta} \\
& \qquad
 + 2
 c_{23}^{2}
  \left(
    s_{\delta} {\rm Re}[\epsilon_{e \mu}^{m}] 
    + c_{\delta} {\rm Im}[\epsilon_{e \mu}^{m}]
  \right)
 \left( \frac{\bar{a}}{4 E_{\nu}} L \right)
  \left(
   \frac{\delta m_{31}^{2}}{4 E_{\nu}} L 
  \right)^{2}
\label{eq:HighEnergy_nuenumu_EMu_m_ImCosDelta} \\
& \qquad
 + 2 
   \left(
    c_{\delta} {\rm Re} [\epsilon_{e \mu}^{m}] 
    - s_{\delta} {\rm Im} [\epsilon_{e \mu}^{m}]    
   \right) \nonumber \\
& \qquad \quad
 \times
   \Biggl\{
    1 - \frac{1}{3} \left( \frac{\bar{a}}{4 E_{\nu}} L \right)^{2}
    + \left(c_{23}^{2} s_{13}^{2} + \frac{2}{3} s_{23}^{2} 
            c_{2 \times 13}
      \right)
      \left( \frac{\bar{a}}{4 E_{\nu}} L \right)
      \left( \frac{\delta m_{31}^{2}}{4 E_{\nu}} L \right)    
   \Biggr\}  
    \left( \frac{\bar{a}}{4 E_{\nu}} L \right)
    \left( \frac{\delta m_{31}^{2}}{4 E_{\nu}} L \right)
\Biggr] 
\label{eq:HighEnergy_nuenumu_EMu_m_ReCosDelta}.
\qquad 
\end{align}
\end{subequations} 
\end{widetext}
Contribution for the transition probability labeled 
\eqref{eq:HighEnergy_nuenumu_EMu_s_ImCosDelta}, 
\eqref{eq:HighEnergy_nuenumu_EMu_s_ReCosDelta} and 
\eqref{eq:HighEnergy_nuenumu_EMu_m_ReCosDelta}
depends on $1/E_{\mu}$.
Consequently, the sensitivities to the terms must be robust against 
the uncertainties of the theoretical parameters 
since in the high energy region they can be
distinguished from the main oscillation part by observing the 
energy dependence.
The claims mentioned above are confirmed numerically by
Fig.\ref{fig:nuenumu_EMu_U0} and \ref{fig:nuenumu_EMu_U10}.  
By comparison of these graphs, 
we can see that the sensitivities to observe the
contribution of \eqref{eq:HighEnergy_nuenumu_EMu_s_ImCosDelta}, 
\eqref{eq:HighEnergy_nuenumu_EMu_s_ReCosDelta} and 
\eqref{eq:HighEnergy_nuenumu_EMu_m_ReCosDelta}
do not suffer
from the uncertainties.\footnote{ To clarify the statement here we adopt
large values for $\epsilon^{s,m}_{e\mu}$. The value of $\epsilon_{e
\mu}$ in Fig.\ref{fig:nuenumu_EMu_U0} and \ref{fig:nuenumu_EMu_U10} are
too large to avoid the current experimental bound. As we said first,
$\epsilon_{e \mu}^{s,m}$ are bound so strictly that these terms can not
be observed.}
Incidentally, 
we note that 
though the uncertainties wreck the sensitivity
to \eqref{eq:HighEnergy_nuenumu_EMu_m_ImCosDelta} 
since it is proportional to $1/E_{\mu}^{2}$, 
the $\epsilon_{e \mu}^{m}$ second order term brings constant 
contribution for energy and this signal does not vanish.

\subsubsection{$\nu_{\mu} \rightarrow \nu_{\mu} $ channel in a neutrino
factory}

\begin{figure*}[th]
\unitlength=1cm
\begin{picture}(16,4)
\put(1.75,-0.4){$E_{\mu}$[GeV]}
\put(5.8,-0.4){$E_{\mu}$[GeV]}
\put(9.8,-0.4){$E_{\mu}$[GeV]}
\put(13.85,-0.4){$E_{\mu}$[GeV]}
\put(-0.5,1.4){\rotatebox{90}{$L$[km]}}
\includegraphics[width=16cm]{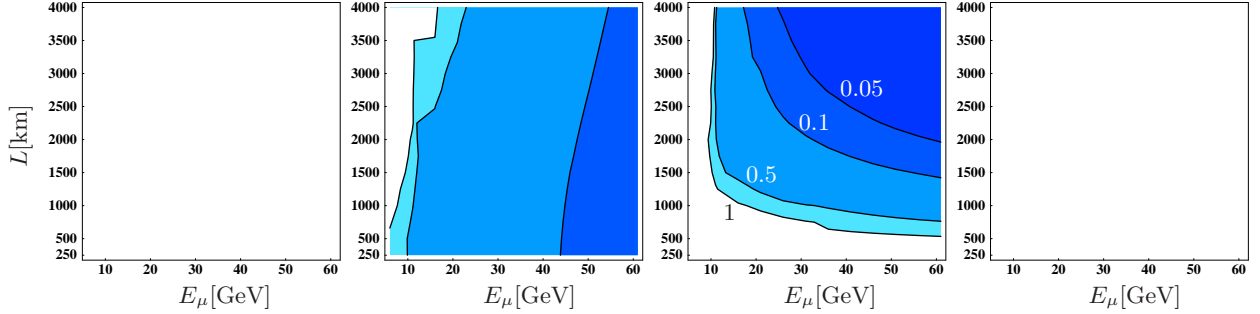}
\put(-7,0.7){1}
\color[rgb]{1,1,1}
\put(-6.7,1.2){0.5}
\put(-6,1.9){0.1}
\put(-5.45,2.35){0.05}
\color[rgb]{0,0,0}
\end{picture}
\vspace{0.3cm}
\caption{Contour plots of the required $N_{\mu} M_{\rm det}$ to observe 
the new physics effects concerning $\epsilon_{\mu \tau}^{s,m}$ 
at 90\% C.L.
in $\nu_{\mu} \rightarrow \nu_{\mu}$ channel.
All theoretical parameters are assumed to have 10\% uncertainty.
From left to right, each plot corresponds to 
$(\epsilon_{\mu \tau}^{s}, \epsilon_{\mu \tau}^{m}) = 
 (3.0 \times 10^{-3},   0)                   $, 
$(3.0 \times 10^{-3} i, 0)                   $, 
$(0,                    3.0 \times 10^{-3})  $, 
$(0,                    3.0 \times 10^{-3} i)$.}
\label{fig:numunumu_MuTau}
\end{figure*}

Only the effect that depend on $\epsilon_{\mu \tau}^{s, m}$
will be large enough to be observed in 
$\nu_{\mu} \rightarrow \nu_{\mu}$ disappearance channel.
As we show in the next section, 
this quantity is not strongly bound by the charged lepton processes.
The analytic expressions for the terms concerning $\epsilon_{\mu \tau}^{s,
m}$ (see the Appendix B) indicate two facts: 
(1) this channel sensitive only to the real part of 
$\epsilon_{\mu \tau}^{m}$, and 
(2) the effect that comes from the real part of $\epsilon_{\mu
\tau}^{s}$ will be small in the assumed parameter region.
In addition, it shows that the terms depending on $\epsilon_{\mu \tau}^{s,m}$ 
are hard to be absorbed by the uncertainty of the theoretical parameters.
Note that these terms do not depend on the CP phase $\delta$, that is,
it is expected that we can get information on the phases of 
$\epsilon_{\mu \tau}^{s,m}$.
The sensitivity plots calculated numerically are shown in
Fig.\ref{fig:numunumu_MuTau}.
They behave as expected by the analytic expressions.
The uncertainties of the theoretical parameters 
do not affect the observability since 
the terms that depend on $1/E_{\mu}$ do not vanish in the high energy
region. 
The sensitivities depend strongly on the phase of $\epsilon_{\mu \tau}^{s,m}$
but do not depend on $\delta$.
We can directly know the phase of the lepton-flavor violation
process without care of $\delta$.

\subsubsection{Channels with $\tau$ and $e$ observation in
a neutrino factory}

The technologies for tau observation in $\nu_{\tau}$ detection
\cite{OPERA}
and the charge identification of electron to distinguish $\nu_{e}$ with 
$\bar{\nu}_{e}$ \cite{Rubbia} are under R \& D.
If it is possible to observe these particles clearly, what can we get ?

\begin{figure*}[th]
\unitlength=1cm
\begin{picture}(16,4)
\put(1.75,-0.4){$E_{\mu}$[GeV]}
\put(5.8,-0.4){$E_{\mu}$[GeV]}
\put(9.8,-0.4){$E_{\mu}$[GeV]}
\put(13.85,-0.4){$E_{\mu}$[GeV]}
\put(-0.5,1.4){\rotatebox{90}{$L$[km]}}
\includegraphics[width=16cm]{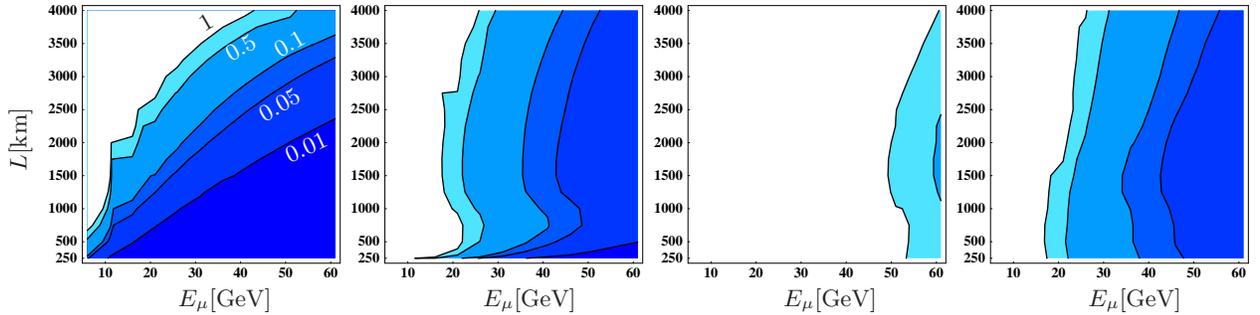}
\put(-14,3.25){\rotatebox{30}{1}}
\color[rgb]{1,1,1}
\put(-13.65,2.85){\rotatebox{30}{0.5}}
\put(-13,2.9){\rotatebox{30}{0.1}}
\put(-13.2,2.05){\rotatebox{30}{0.05}}
\put(-12.85,1.5){\rotatebox{25}{0.01}}
\color[rgb]{0,0,0}
\end{picture}
\vspace{0.3cm}
\caption{Contour plots of the required $N_{\mu} M_{\rm det}$ to observe 
the new physics effects concerning $\epsilon_{e \tau}^{s,m}$ 
at 90\% C.L.
in $\nu_{e} \rightarrow \nu_{\tau}$ channel.
All theoretical parameters are assumed to have 10\% uncertainty.
From left to right, each plot corresponds to 
$(\epsilon_{e \tau}^{s}, \epsilon_{e \tau}^{m}) = 
 (3.0 \times 10^{-3},   0)                   $, 
$(3.0 \times 10^{-3} i, 0)                   $, 
$(0,                    3.0 \times 10^{-3})  $, 
$(0,                    3.0 \times 10^{-3} i)$.}
\label{fig:nuenutau_ETau}
\end{figure*}

\begin{figure*}[th]
\unitlength=1cm
\begin{picture}(16,4)
\put(1.75,-0.4){$E_{\mu}$[GeV]}
\put(5.8,-0.4){$E_{\mu}$[GeV]}
\put(9.8,-0.4){$E_{\mu}$[GeV]}
\put(13.85,-0.4){$E_{\mu}$[GeV]}
\put(-0.5,1.4){\rotatebox{90}{$L$[km]}}
\includegraphics[width=16cm]{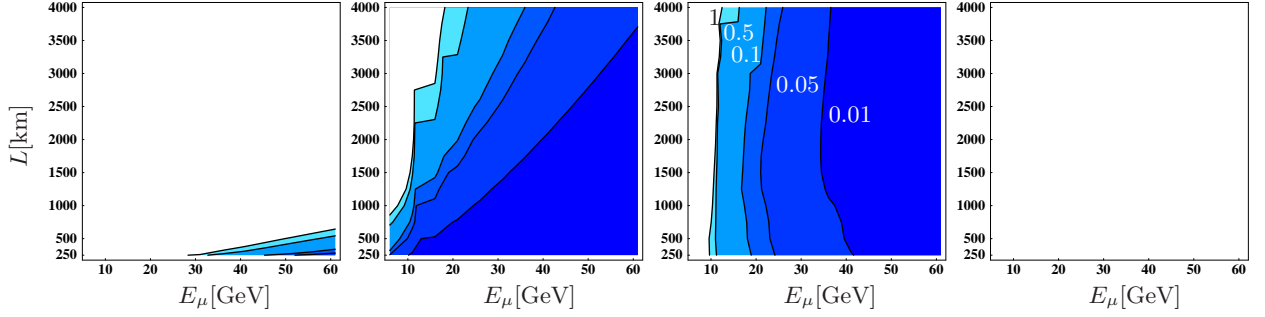}
\put(-7.2,3.3){1}
\color[rgb]{1,1,1}
\put(-7,3.1){0.5}
\put(-6.9,2.8){0.1}
\put(-6.3,2.4){0.05}
\put(-5.6,2){0.01}
\color[rgb]{0,0,0}
\end{picture}
\vspace{0.3cm}
\caption{Same as Fig.\ref{fig:numunumu_MuTau}, 
but $\nu_{\mu} \rightarrow \nu_{\tau}$ channel.}
\label{fig:numunutau_MuTau}
\end{figure*}

\begin{figure*}[th]
\unitlength=1cm
\begin{picture}(16,4)
\put(1.75,-0.4){$E_{\mu}$[GeV]}
\put(5.8,-0.4){$E_{\mu}$[GeV]}
\put(9.8,-0.4){$E_{\mu}$[GeV]}
\put(13.85,-0.4){$E_{\mu}$[GeV]}
\put(-0.5,1.4){\rotatebox{90}{$L$[km]}}
\includegraphics[width=16cm]{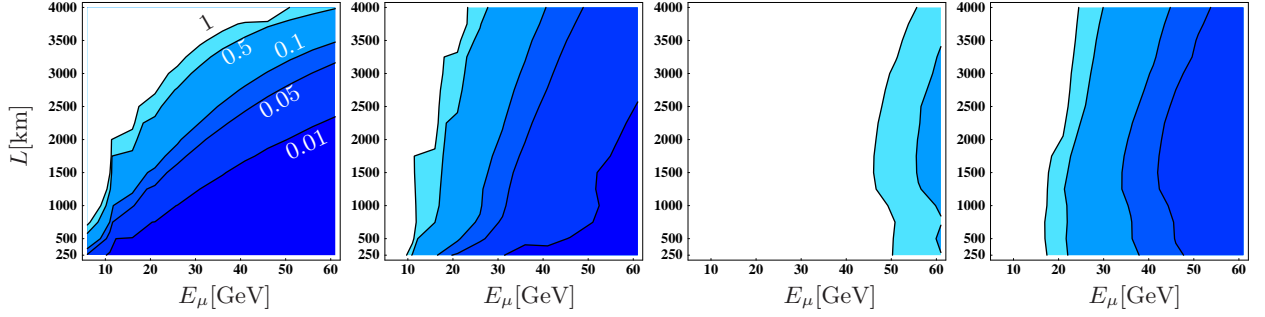}
\put(-14,3.2){\rotatebox{30}{1}}
\color[rgb]{1,1,1}
\put(-13.7,2.75){\rotatebox{30}{0.5}}
\put(-13,2.85){\rotatebox{25}{0.1}}
\put(-13.2,2.05){\rotatebox{30}{0.05}}
\put(-12.85,1.5){\rotatebox{25}{0.01}}
\color[rgb]{0,0,0}
\end{picture}
\vspace{0.3cm}
\caption{Contour plots of the required $N_{\mu} M_{\rm det}$ to observe 
the new physics effects concerning $\epsilon_{\mu e}^s$ and
$\epsilon^m_{e\mu}$ at 90\% C.L.
in $\nu_{\mu} \rightarrow \nu_{e}$ channel.
All theoretical parameters are assumed to have 10\% uncertainty.
From left to right, each plot corresponds to 
$(\epsilon_{\mu e}^{s}, \epsilon_{e \mu}^{m}) = 
 (3.0 \times 10^{-3},   0)                   $, 
$(3.0 \times 10^{-3} i, 0)                   $, 
$(0,                    3.0 \times 10^{-3})  $, 
$(0,                    3.0 \times 10^{-3} i)$.}
\label{fig:numunue_EMu}
\end{figure*}

In $\nu_{e} \rightarrow \nu_{\tau}$ channel,
we can explore $\epsilon_{e \tau}^{s,m}$ (see Fig.\ref{fig:nuenutau_ETau}). 
The uncertainties of the theoretical parameters 
will not disturb the observability.
In $\nu_{\mu} \rightarrow \nu_{\tau}$ channel, 
all we can observe is only the effect of $\epsilon_{\mu \tau}^{s,m}$ 
(see Fig.\ref{fig:numunutau_MuTau}).
In comparison with $\nu_{\mu} \rightarrow \nu_{\mu}$ channel,
we will not have so much benefit 
in terms of sensitivities to the magnitude of $\epsilon$.%
\footnote{Including systematic errors we will have much benefit.}
However in this channel unlike $\nu_{\mu} \rightarrow \nu_{\mu}$ channel,
the observable must depend on the combination of the 
$\epsilon_{\mu \tau}^{s,m}$'s phase and the CP phase $\delta$, that is, 
the observation in these two channels are qualitatively different.
In $\nu_{\mu} \rightarrow \nu_{e} $ channel, 
we can search the effect of not $\epsilon^s_{e \mu}$ but 
$\epsilon^s_{\mu e}$ at the muon decay process 
and the effect of $\epsilon^m_{e\mu}$ at the propagation process
just like $\nu_{e} \rightarrow \nu_{\mu}$ channel 
(see Fig.\ref{fig:numunue_EMu}).
In $\nu_{e} \rightarrow \nu_{e}$ disappearance channel, 
oscillation effects themselves are much smaller 
than the no-oscillation signal.
Though some effects of new physics give the different energy dependence 
from the main oscillation term,
we will not be able to get any information for oscillation-enhanced 
new physics.

\subsection{$(V-A)(V+A)$ type interactions}

\begin{figure*}[th]
\unitlength=1cm
\begin{picture}(12,4)
\put(1.8,-0.4){$E_{\mu}$[GeV]}
\put(5.8,-0.4){$E_{\mu}$[GeV]}
\put(9.85,-0.4){$E_{\mu}$[GeV]}
\put(-0.5,1.4){\rotatebox{90}{$L$[km]}}
\includegraphics[width=12cm]{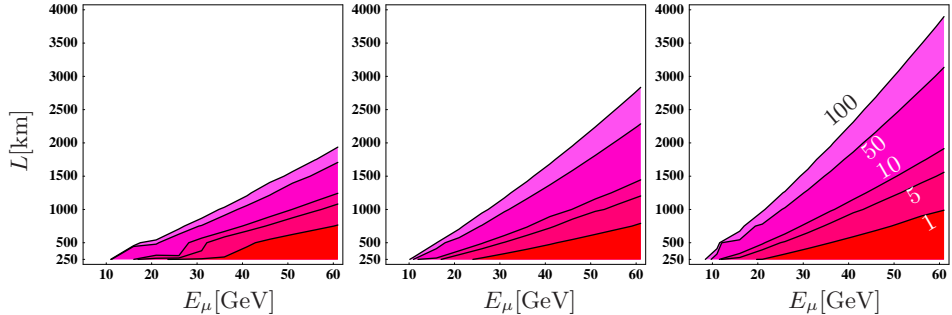}
\put(-1.7,2){\rotatebox{40}{100}}
\color[rgb]{1,1,1}
\put(-1.2,1.55){\rotatebox{40}{50}}
\put(-1,1.3){\rotatebox{30}{10}}
\put(-0.6,0.97){\rotatebox{30}{5}}
\put(-0.4,0.6){\rotatebox{30}{1}}
\color[rgb]{0,0,0}
\end{picture}
\vspace{0.3cm}
\caption{Contour plots of the required $N_{\mu} M_{\rm det}$
 to observe the $(V-A)(V+A)$ type new interaction, 
${\epsilon'}_{\mu \tau}^{s} = 3.0 \times 10^{-3} i$,
at 90\% C.L.
in $\nu_{\mu} \rightarrow \nu_{\tau}$ channel.
Contours mean 1, 5, 10, 50, 100$\times 10^{21} \cdot 100{\rm kt}$.
Here, the uncertainties of the theoretical parameters are not considered. 
From left to right, each plot corresponds to $\mathcal{P_{\mu}} =$0.99,
 0, $-0.99$.}
\label{fig:V+A}
\end{figure*}

The signals of $(V-A)(V+A)$ type interactions are hard to be observed because
they are suppressed by the factor, 
$m_{e}/ m_{\mu}$.
The interference term between the leading term and $(V-A)(V+A)$ type ones 
has different polarization dependence from that of the leading contribution 
unlike $(V-A)(V-A)$ type new interaction. 
If we can make good use of this fact,
then we may be able to expect to gain somewhat better sensitivity.
As we saw in eqs.\eqref{eq:cross-amp} and \eqref{eq:cross-amp-numu-forward}, 
the utilization of the muon polarization works only 
in $\nu_{\mu}\rightarrow \nu_{\alpha}$. 
Figure.\ref{fig:V+A} describes the sensitivity to 
${\epsilon'}_{\mu \tau}^{s} ( \equiv \lambda' / G_{F})$ 
in $\nu_{\mu} \rightarrow \nu_{\tau}$ channel.
We can gain a little, but have no advantage over a direct search against
our expectation.

\subsection{$\nu_{\mu} \rightarrow \nu_{e, \mu, \tau} $ channel in
an upgraded conventional beam}

In an experiment using a conventional beam,
there are some different points from the case of a neutrino factory.
Since we consider two body decay of $\pi$'s, the created charged lepton
and the neutrino have the fixed helicity whatever new physics is.
It means that we do not have to worry the type of a new interaction.
We can always parameterize the effect of new physics concerning $\pi$
decay by $\epsilon^s_{\mu e}$ and $\epsilon^s_{\mu\tau}$.
When we consider a new source interaction, new interactions 
between neutrino and {\it nucleon} must be introduced not only in matter
but also in detector. 
However to consider such effects we have to treat 
complicated hadronic processes, 
so we do not consider the relation between these two effects.
The contribution to the source and  the matter effect is not always generated 
by the same new interactions.
Hence, we can suppose that different magnitude interaction in each process. 

\begin{figure*}[th]
\unitlength=1cm
\begin{picture}(8,4)
\put(1.75,-0.4){$E_{\nu}$[GeV]}
\put(5.8,-0.4){$E_{\nu}$[GeV]}
\put(-0.5,1.4){\rotatebox{90}{$L$[km]}}
\includegraphics[width=8cm]{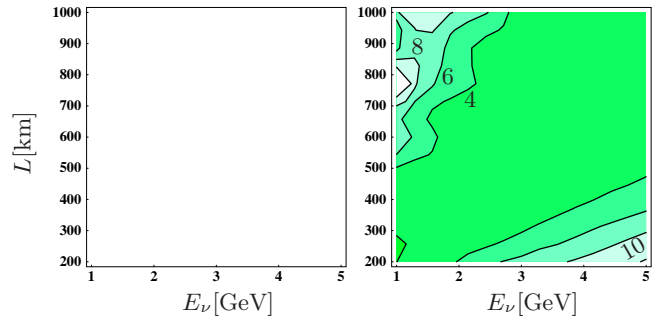}
\put(-2.5,2.3){4}
\put(-2.8,2.6){6}
\put(-3.2,3){8}
\put(-0.45,0.25){\rotatebox{30}{10}}
\end{picture}
\vspace{0.3cm}
\caption{Contour plots of the required no-oscillated neutrino number, $N_{\nu}$, 
to observe the new physics effects concerning $\epsilon_{\mu \tau}^{s}$ 
at 90\% C.L.
in $\nu_{\mu} \rightarrow \nu_{\mu}$ channel using an upgraded
conventional beam.
All theoretical parameters are assumed to have 10\% uncertainty.
From left to right, each plot corresponds to 
$\epsilon_{\mu \tau}^{s} = 
 3.0 \times 10^{-3}$, 
$3.0 \times 10^{-3} i $.}
\label{fig:conv_numunumu_MuTau}
\end{figure*}

\begin{figure*}[th]
\unitlength=1cm
\begin{picture}(8,4)
\put(1.75,-0.4){$E_{\nu}$[GeV]}
\put(5.8,-0.4){$E_{\nu}$[GeV]}
\put(-0.5,1.4){\rotatebox{90}{$L$[km]}}
\includegraphics[width=8cm]{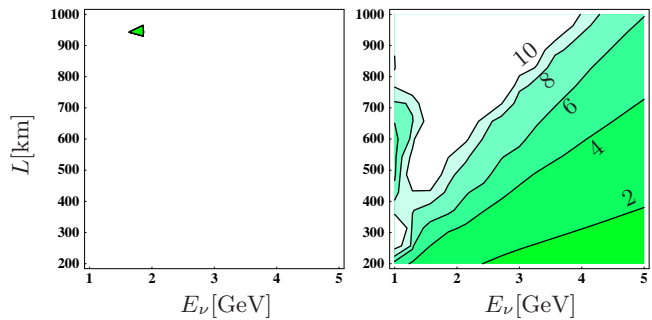}
\put(-1.85,2.85){\rotatebox{40}{10}}
\put(-1.5,2.55){\rotatebox{40}{8}}
\put(-1.2,2.2){\rotatebox{40}{6}}
\put(-0.85,1.65){\rotatebox{40}{4}}
\put(-0.4,1){\rotatebox{25}{2}}
\end{picture}
\vspace{0.3cm}
\caption{Same as Fig.\ref{fig:conv_numunumu_MuTau}, 
but $\nu_{\mu} \rightarrow \nu_{\tau}$ channel.
}
\label{fig:conv_numunutau_MuTau}
\end{figure*}

As it became clear in analysis for a neutrino factory, 
all we can probe is the contribution of $\epsilon_{\mu e}^{s}$, 
$\epsilon_{e \mu}^{m}$ in $\nu_{\mu} \rightarrow \nu_{e}$ channel 
and that of $\epsilon_{\mu \tau}^{s,m}$ in 
$\nu_{\mu} \rightarrow \nu_{\mu, \tau}$ channel. 
However, new
interactions in matter can not be observed in the energy and the baseline
regions that we assume here and hence we do not study their effect.  
The numerical results for the necessary number of no-oscillated neutrinos 
$N_{\nu}$ are shown in Fig.\ref{fig:conv_numunue_EMu},
Fig.\ref{fig:conv_numunumu_MuTau}, and
Fig.\ref{fig:conv_numunutau_MuTau}.  
The neutrino event number of current proposed
experiments is estimated of  $\mathcal{O} (10^{3} \sim 10^{4})$,
and $\mathcal{O}(10^{2})$ times larger exposure is expected in the next
generation.

Figures \ref{fig:conv_numunue_EMu}, \ref{fig:conv_numunumu_MuTau}
and \ref{fig:conv_numunutau_MuTau}
show that the sensitivity strongly depends on the complex phase
of $\epsilon$'s. 
This fact means that the observations give us information
of the phase in lepton-flavor-violation interactions.  
Depending on models, some interesting issues are revealed.  
In the next section, we discuss models which may give the significant 
$\epsilon$'s. 
As we point out above, $\epsilon$'s for a neutrino factory and a conventional 
beam have
different dependence on new interactions.  Hence, it is important for
new physics search to compare the results of a neutrino factory and that
of a conventional beam.

\begin{figure*}[th]
\unitlength=1cm
\begin{picture}(8,4)
\put(1.75,-0.4){$E_{\nu}$[GeV]}
\put(5.8,-0.4){$E_{\nu}$[GeV]}
\put(-0.5,1.4){\rotatebox{90}{$L$[km]}}
\includegraphics[width=8cm]{conv-MutoE-EpsilonEMu-U10-No2.epsi}
\put(-5.7,2.65){\rotatebox{40}{10}}
\put(-5.2,2.3){\rotatebox{40}{8}}
\put(-5,1.9){\rotatebox{30}{6}}
\put(-4.8,1.5){\rotatebox{30}{4}}
\put(-4.4,0.7){\rotatebox{20}{2}}
\end{picture}
\vspace{0.3cm}
\caption{Contour plots of the required no-oscillated neutrino number, 
$N_{\nu}$, 
to observe the new physics effects concerning $\epsilon_{\mu e}^{s}$ 
at 90\% C.L.
in $\nu_{\mu} \rightarrow \nu_{e}$ channel using an upgraded conventional beam.
All theoretical parameters are assumed to have 10\% uncertainty.
From left to right, each plot corresponds to 
$\epsilon_{\mu e}^{s} = 
 3.0 \times 10^{-3}                   $, 
$3.0 \times 10^{-3} i                   $.
In this energy and baseline region, a experiment does not sensitive 
to new interactions in matter.
In Fig.\ref{fig:conv_numunue_EMu}, Fig.\ref{fig:conv_numunumu_MuTau},
and  Fig.\ref{fig:conv_numunutau_MuTau}, contours mean 2, 4, 6, 8, 10
 $\times 10^{5}$ no-oscillated neutrinos.}
\label{fig:conv_numunue_EMu}
\end{figure*}

\subsection{short summary for numerical analysis}
For a neutrino factory:
\begin{itemize}
 \item In $\nu_{\alpha} \rightarrow \nu_{\beta}$, $(\alpha=e, \mu,$
       $\beta=e, \mu, \tau)$ appearance channel, 
       the observable effects of new $(V-A)(V-A)$
       interactions come only  from $\epsilon_{\alpha \beta}^{s,m}$.
       The others are too small or too vulnerable against the 
       adjustment of the theoretical parameters.
       Note that $\delta$ and $\epsilon$'s phase are correlated.
       Namely the measured values are a certain combination of $\delta$
       and $\epsilon$.

 \item In $\nu_{\mu} \rightarrow \nu_{\mu}$ disappearance channel,
       we can measure $\epsilon_{\mu \tau}^{s,m}$ depending on their phase.
       In other words, the signal includes information of the phase.
       Furthermore, there is no correlation between $\delta$ and
       $\epsilon$, so the measurement tells us directly
       the phase concerning the lepton-flavor violating process.
       In $\nu_{e} \rightarrow \nu_{e}$ disappearance channel, we can
       not get anything for new interactions in the oscillation
       enhanced way.

 \item The $\chi^{2}$ is proportional to $|\epsilon|^{2}$.
       The expected sensitivity is to $|\epsilon| \gtrsim \mathcal{O}
       (10^{-4})$ by using this methodology. 
 
 \item When the situations that new interactions exist not only
       in the source but also in the matter effect are considered, 
       we can easily understand the sensitivity by simply 
       adding each effect.

  \item Oscillation-enhanced effects for the $(V-A)(V+A)$ type interactions are
       strongly suppressed by $m_{e}/m_{\mu}$, 
       so we can not get an advantage over a direct measurement.
\end{itemize}

For an upgraded conventional beam:
\begin{itemize}
 \item We do not have to care the types of new interactions in the source.
       The analyses for the feasibility are similar to that 
       of $(V-A)(V-A)$ type for a neutrino factory.
       In the assumed energy and baseline region, there is no
       sensitivity to the new effect in matter.   

 \item The $\epsilon$'s for a conventional beam have  different dependence
       from those for a neutrino factory on new interactions. Therefore,
       the comparison between two methods makes clear the species of 
       new physics.

\end{itemize}
\begin{center}
\begin{table}[th]
\begin{ruledtabular}
\begin{tabular}{cccc}
 & $\epsilon_{e \mu}^{s,m}$, $(\epsilon_{\mu e}^{s})$
  & $\epsilon_{e \tau}^{s,m}$ 
   & $\epsilon_{\mu \tau}^{s,m}$ \\
\hline
bound from charged lepton processes & $5 \times 10^{-5} $ & --- 
   & --- \\
 \hline \hline
$\nu_{e} \rightarrow \nu_{\mu}$ & \hspace{0.042cm}
                                  {\large $\triangle$}\footnotemark[1] 
                                & \hspace{0.042cm}
                                  {\large $\triangle$}\footnotemark[2]
                                & {\Large $\times$} \\
$\nu_{\mu} \rightarrow \nu_{\mu}$ & {\Large $\times$} 
                                  & {\Large $\times$}
                                  & \hspace{0.042cm}
                                    {\Huge $\circ$}\footnotemark[3]
 \\
 \hline
$\nu_{e} \rightarrow \nu_{\tau}$ & {\Large $\times$} 
                                 & {\Huge $\circ$}
                                 & \hspace{0.042cm}
                                   {\large $\triangle$}\footnotemark[2]\\
$\nu_{\mu} \rightarrow \nu_{\tau}$ & {\Large $\times$} 
                                   & \hspace{0.042cm}
                                     {\large $\triangle$}\footnotemark[2]
                                   & \hspace{0.042cm}
                                     {\Huge $\circ$}\footnotemark[3]
 \\
 \hline
$\nu_{\mu} \rightarrow \nu_{e} $ & \hspace{0.042cm}
                                   {\large $\triangle$}\footnotemark[1] 
                                 & {\Large $\times$}
                                 & {\Large $\times$}\\
$\nu_{e} \rightarrow \nu_{e} $   & {\Large $\times$}
                                 & {\Large $\times$}
                                 & {\Large $\times$}
\end{tabular}
\end{ruledtabular}
\footnotetext[1]{appropriate mode but bound is too strict}
\footnotetext[2]{too vulnerable against the adjustment of the 
theoretical parameters}
\footnotetext[3]{depending on the $\epsilon$'s phase, in other words,
 sensitive to the phase}
\caption{Summary on the feasibility of $(V-A)(V-A)$ type new interactions.}
\end{table}
\end{center}

\section{Indications to exotic decays from various models}

In this section we discuss models which give exotic interactions
interfering with the weak interaction in the neutrino oscillation and
survey to which process those exotic interactions contribute.  We should
consider models which give an explanation for the smallness of the
neutrino masses and the lepton mixings. 
In those models we can expect that flavor violating processes are induced,
and
since to explain the neutrino masses and the lepton mixings
we need to introduce sources of flavor violations.

There are well-known two types of models which explain the smallness of
the neutrino masses and the lepton mixings.  One is a seesaw 
type \cite{seasaw} and the other is a radiative type \cite{zee}.  
These models induce
other flavor violating effects generally.  Such other flavor violating
contribution could be large enough to give observable effects for neutrino
oscillation experiments through the interfering effects. We now consider
these two types of models concerning mainly on supersymmetric models and
study how large the exotic effective interactions are.

\subsection{Supersymmetric models with right-handed neutrinos \label{sec:5-1}}
Among the promising extensions of the standard model, a supersymmetric
standard (or grand unified) model with right-handed neutrinos is
often considered. In this class of models, if the gravity-induced
supersymmetry breaking is employed, through the renormalization effect,
the large flavor violating slepton masses are induced \cite{Bor}, even if
at the cutoff scale there is no flavor violating slepton mass.

\begin{figure}[h]
\center
\epsfig{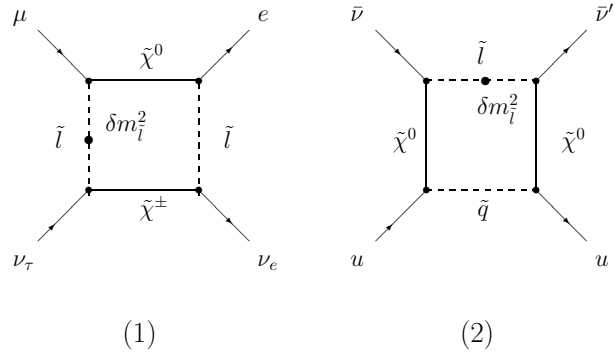}
\caption{(1) An Example of one-loop diagrams contributing to the $\mu$ decay
  and the matter effect.
(2) An Example of tree diagrams contributing to the $\pi$ decay, the
 matter effect and the detection process. 
Where $\delta m^{2}_{\tilde{l}}$ denotes an off-diagonal element of the
mass matrix for sleptons which violates the lepton flavor.}
\label{fig:box}
\end{figure}
Due to the lepton flavor violating masses,
 one-loop box diagrams constructed by propagations of superpartners 
induce exotic effective $(V-A)(V-A)$ interactions, which contribute to
the decay side, the matter effect and the detection side all together.  
One of effective interactions contributing to the decay side in oscillation
experiments based on a muon beam is drawn in Fig\ref{fig:box}-(1). 
This gives $\epsilon_{\mu\tau}^s$. There is a constraint on the magnitude of
Fig\ref{fig:box}-(1) from the lepton flavor violating decays of $\tau$ such
as $\tau\rightarrow\mu\gamma$ and $\tau\rightarrow\mu e e$. 
However, these constraints are not so strict 
and $\epsilon_{\mu\tau}^s$ can be of ${\mathcal O}(10^{-3})$.  
This type of the effective interaction also induces the matter
effect similarly replacing $\nu_e$ with $e$ and $\mu$ with $\nu_\mu$
 respectively, and also 
$\epsilon^m_{\mu\tau}$ can be of ${\mathcal O}(10^{-3})$. 
By changing the
external legs in Fig \ref{fig:box}-(1), 
we can draw similar diagrams for four-lepton couplings 
which violate the lepton flavor,
and using them we can estimate how large 
values the $\epsilon^{s,m}$'s are in each model and experimentally. 
They can contribute to the $\mu$ decay and the matter effects.

In addition, there are effective interactions generated by one-loop
diagrams including squark propagators as Fig\ref{fig:box}-(2).
These kinds of diagrams can contribute to the detection processes 
and the matter effects.
Furthermore they can contribute to the $\pi$ decay. 

These two kinds of diagrams affect the oscillation in a neutrino factory
and a conventional beam differently. 
By comparing the results of these two 
experiments, we may have some information on the scalar masses. For
example, if the squark masses are much heavier than the slepton
masses, then
only the figures of the type Fig\ref{fig:box}-(1) contribute to the
oscillation phenomena, and hence there is a difference between
``oscillation probabilities'' in these two experiments. 
On the other hand
if the magnitudes of the scalar masses are comparable, these two 
experiments are affected from new physics in the decay side, 
the matter effect and the detection side similarly.

It is important that we can get information on the
phases of $\epsilon$'s. Their phases arise from the phases of the flavor
changing slepton masses, whose origins are the Yukawa couplings for
$\nu_L$ and $\nu_R$ at the high energy scale. We may be able to acquire 
information on the phases of Dirac-Yukawa couplings.

\subsection{Models with radiatively induced neutrino masses\label{sec:5-2}}
\begin{figure}[h]
\center
\epsfig{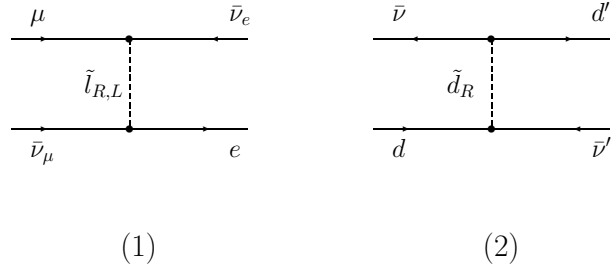}
\caption{(1) Examples of tree diagrams contributing to the $\mu$ decay
 and the matter effect.
(2) An Example of tree diagrams contributing to the $\pi$ decay, the
 matter effect and the detection process. 
The first diagram (1) induces both $(V-A)(V-A)$ type and $(V-A)(V+A)$ type interactions
for the muon decay and the matter effect.
The second diagram (2) induces $(V-A)(V+A)$ type interactions
only, which contribute to the matter effect.}
\label{fig:tree}
\end{figure}

There are many kinds of models in which the neutrino masses and the
lepton mixings are radiatively induced.  Among them we consider 
supersymmetric models with R-parity violating terms. A most general
superpotential which breaks R-parity is
\begin{eqnarray}
W_{\rm RPV} = \lambda_{\alpha\beta\gamma}
 L_{\alpha}L_{\beta}E^{c}_{\gamma}
 +\lambda'_{\alpha\beta\gamma}
L_{\alpha}Q_{\beta}D^{c}_{\gamma} 
+\lambda''_{\alpha\beta\gamma} U^{c}_{\alpha}D^{c}_{\beta}D^{c}_{\gamma} +
\mu'_{\alpha}L_{\alpha}H_{U\alpha}, \label{eq:RPV-SP}
\end{eqnarray}
 where $\lambda_{\alpha\beta\gamma} =
-\lambda_{\beta\alpha\gamma},\lambda''_
{\alpha\beta\gamma} = -\lambda''_{\alpha\gamma\beta}$. $L_\alpha$, 
$E^c_\alpha$,
$Q_\alpha$, $U_\alpha$,  $D^c_\alpha$
and $H_{U\alpha}$
 are the superfields corresponding to the lepton
doublets, the right-handed charged leptons, the quark doublets, 
the right-handed up-type-quarks, the down-type quarks and the up-type
Higgs doublet respectively.  
Three trilinear terms
are allowed if the R-parity is broken generally. 
However there is no reason 
why all R-parity violating terms should be included at the same
time. For example, we forbid interactions proportional to
$\lambda''_{\alpha\beta\gamma}$ by imposing the baryon-number
conservation 
to avoid rapid proton decay. 
To induce small masses for neutrinos, the first term
or the second term in eq.\eqref{eq:RPV-SP} must be included. For example the
first term in eq.\eqref{eq:RPV-SP} has interactions described as
\begin{eqnarray}
&& {\mathcal L}= \lambda_{\alpha\beta\gamma}
( \bar{\nu}^{c}_{\alpha L} e_{\beta L}\tilde{e}_{\gamma R}^{*} +
\bar{e}_{\gamma R} \nu_{\alpha L}\tilde{e}_{\beta L} + 
\bar{e}_{\gamma R}e_{\beta L}\tilde{\nu}_{\alpha L}  
-( \alpha \leftrightarrow \beta )) + {\rm H.c}.  \label{eq:RPV-1}
\end{eqnarray}
These interactions work in a same way as those of the Zee model.\cite{zee}
Namely, sleptons interact with charged leptons and neutrinos as the Zee
boson, and the small neutrino masses and the lepton mixings are induced
radiatively.  Such interactions induce $(V-A)(V-A)$ type and $(V-A)(V+A)$
type effective four-lepton interactions which violate the lepton flavor
(see Fig.\ref{fig:tree}). These affect oscillation experiments through
the interference with the weak interaction. Since there are only
four-lepton type effective interactions, we would be able to
see the difference between ``oscillation probabilities'' 
in two different experiments.
  In the case of a
neutrino factory, they affect both the decay side and the matter effect
all together, while in the case of a conventional beam they affect only
the matter effect. From these kinds of Fig.\ref{fig:tree}-(1)
$\epsilon^{s,m}_{\mu\tau}$ and $\epsilon^s_{e\tau}$ are induced.
In this case the magnitudes of $\epsilon$'s could be larger than
the case in the previous subsection.

Interactions generated by the second term in eq.{\eqref{eq:RPV-SP}},
are described as
\begin{eqnarray}
&&{\mathcal L} =\lambda_{\alpha\beta\gamma}'V^{KM}_{\beta\delta}
( \bar{\nu}^{c}_{\alpha L}
d_{\delta L}\tilde{d}_{\gamma R}^{*} + \bar{d}_{\gamma R} \nu_{\alpha
  L}\tilde{d}_{\delta L} +
\bar{d}_{\gamma R} d_{\delta L}\tilde{\nu}_{\alpha L} ) \nonumber \\ &&~~~~~~+
\lambda_{\alpha\beta\gamma}'( \bar{u}^{c}_{\beta L} e_{\alpha L}
\tilde{d}_{\gamma R}^{*} + \bar{d}_{\gamma R}
e_{\alpha L}\tilde{u}_{\beta L} + \bar{d}_{\gamma R}u_{\beta L}
\tilde{e}_{\alpha L} )   + {\rm H.c.},
\label{eq:RPV-2}
\end{eqnarray}
where $V^{KM}$ is the mixing matrix for the quark sector.
In this case squarks work as the Zee boson similarly. 
These interactions induce effective four-Fermi interactions 
with the lepton flavor violation, 
as drawn in Fig.\ref{fig:tree}-(2).  
They affect the matter effect and
the detection process in both experiments based on a muon beam and on
a conventional beam.
For a conventional beam, there is an additional interference in the decay side.
Therefore if the R-parity violating term is limited only to the second  
term  in
eq.{\eqref{eq:RPV-SP}}, then we also expect to  be  able to 
make sure the existence of such terms by comparison with the
 two different experiments. 

\subsection{
General properties of the effective four-lepton interactions
and order estimation for their couplings.}

Without assuming physics that gives the effective four-lepton
interactions containing two left-handed neutrinos and two charged
leptons with the lepton flavor violation,
we can classify two types of such interactions from Lorentz invariance as 
the $(V-A)(V-A)$ type and the $(V-A)(V+A)$ type generally.

The flavor violating four-lepton couplings with the $(V-A)(V-A)$ type 
are classified into two categories: SU(2)$_L$ singlet type,
\begin{eqnarray}
&& h_{\alpha\beta\gamma\delta} (\bar{l}_\alpha C \bar l_\beta)
(l_\gamma C^\dagger l_\delta)
\label{eq:singlettype}
\\
&&= -\frac{1}{2}h_{\alpha\beta\gamma\delta}
(\bar{ l}_\alpha \gamma^\mu l_\delta)
(\bar{l}_\beta \gamma_\mu l_\gamma)
\nonumber \\
&&= \frac{1}{2}h_{\alpha\beta\gamma\delta}
 h_{\alpha\beta\gamma\delta} 
 (\bar{e}_{\alpha L} C
 \bar{\nu}_{\beta L} -\bar{\nu}_{\alpha L} C  \bar{e}_{\beta L})
({e}_{\gamma L} C^{\dagger}
 {\nu}_{\delta L} - \nu_{\gamma L} C^{\dagger}  {e}_{\delta L}) ,
\nonumber
\end{eqnarray}
and triplet type,
\begin{eqnarray}
&& g_{\alpha\beta\gamma\delta} (\bar{l}_\alpha \tau^a C \bar l_\beta)
(l_\gamma C^\dagger \tau^a l_\delta)
\label{eq:triplettype}\\
&&= -\frac{1}{2}g_{\alpha\beta\gamma\delta}
(\bar{ l}_\alpha \tau^a \gamma^\mu l_\delta)
(\bar{l}_\beta \tau^a \gamma_\mu l_\gamma) \nonumber \\
&&= g_{\alpha\beta\gamma\delta} \left\{
(\bar{e}_{\alpha L} C \bar{\nu}_{\beta L} + \bar{\nu}_{\alpha L} C \bar{ e}_{\beta L})
(e_{\gamma L} C^{\dagger} \nu_{\delta L} 
+  \nu_{\gamma L} C^{\dagger}  e_{\delta L}) \right.  \nonumber \\
&&~~~~ \left. +2 (\bar{\nu}_{\alpha L}  C \bar{\nu}_{\beta L})
(\nu_{\gamma L} C^\dagger \nu_{\delta L}) +
2 (\bar{e}_{\alpha L}  C \bar{e}_{\beta L})
(e_{\gamma L} C^\dagger e_{\delta L}) \right\}.\nonumber
\end{eqnarray}
Here $\l_{\alpha}$ is a lepton doublet with flavor $\alpha$ and 
$\tau^a$ is a Pauli matrix.
Due to the SU(2)$_L$ invariance coupling constants must satisfy 
$h_{\alpha\beta\gamma\delta}=-h_{\beta\alpha\gamma\delta}=-
h_{\alpha\beta\delta\gamma}$ and $g_{\alpha\beta\gamma\delta}=
g_{\beta\alpha\gamma\delta}=g_{\rho\sigma\delta\gamma}$.
Above two interactions are described as the effective interactions
induced by exchange of not only vector particles but also scalar particles. 
The former type is induced, for example, 
if there is a coupling of the form 
\begin{eqnarray}
 \bar{ l} C \bar{l} \phi_{\rm S} +{\rm H.c.},
\label{eq:zeetype}
\end{eqnarray}
where $\phi_{\rm S}$ is an SU(2)$_L$ singlet.
The latter is induced by exchange of an SU(2)$_L$ triplet $\phi^{a}_{\rm
T}$,
if the coupling 
\begin{eqnarray}
 \bar{ l}\tau^aC\bar{l}\phi^a_{\rm T} + {\rm H.c.},
\label{eq:triplet}
\end{eqnarray}
exists.

On the other hand, the flavor violating four-lepton couplings with $(V-A)(V+A)$
 type have only one category: SU(2)$_L$ doublet type,
\begin{eqnarray}
&& f_{\alpha\beta\gamma\delta} (\bar{l}_\alpha  e_{\beta R})
(\bar{e}_{\gamma R} {\l}_\delta)
\label{eq:doublettype}
\\
&&= -\frac{1}{2} f_{\alpha\beta\gamma\delta} 
(\bar{l}_\alpha \gamma^{\mu}  {l}_\delta )
(\bar{e}_{\gamma} \gamma_{\mu} P_{R} e_\beta)\nonumber\\
&&= f_{\alpha\beta\gamma\delta} \left\{
( \bar{e}_{\alpha L}  e_{\beta R})(\bar{e}_{\gamma R} {e}_{\delta L})
+ ( \bar{\nu}_{\alpha L}  e_{\beta R})(\bar{e}_{\gamma R}
{\nu}_{\delta L}) \right\}.\nonumber
\end{eqnarray}
This type is induced by the exchange of a scalar particle belonging to an
SU(2)$_{L}$ doublet.\footnote{ This type of an interaction is also
  induced by a tensor particle 
exchange.}
Suppose that there is a coupling of the form 
\begin{eqnarray}
 \bar{ l}e\phi_{\rm D} + {\rm H.c.},
\label{eq:triplet}
\end{eqnarray}
where $\phi_{\rm D}$ is an SU(2)$_{L}$ doublet scalar, the interaction 
eq.\eqref{eq:doublettype} is induced.

In general the constraint on the magnitude of the singlet type effective
coupling $h_{\alpha\beta\gamma\delta}$ is rather weak. 
A radiative model often contains such a singlet scalar.
Indeed models in section \ref{sec:5-2} contain such SU(2)$_{L}$ singlet 
scalars (see Fig.\ref{fig:tree}-(1) with $\tilde{l}_{R}$).
As for the $\mu$ decays
which contribute to the interference phenomena,
replacing $\alpha$, $\beta$, $\gamma$ and $\delta$ 
in $h_{\alpha\beta\gamma\delta}$ 
with $e$, $\mu$,
$e$ and  $\tau$ we get $\epsilon^{s,m}_{\mu\tau}$ and 
replacing $\alpha$, $\beta$, $\gamma$ and $\delta$ 
in $h_{\alpha\beta\gamma\delta}$
with $e$, $\mu$, $\mu$ and $\tau$ we get 
$\epsilon^{s}_{e\tau}$. 
In this case there is no
constraint from $\tau\rightarrow lll$ decays. 
Also constraints from
$\tau\rightarrow\mu\gamma$ (for the former) and $\tau\rightarrow e\gamma$
(for the latter) are not severe.\cite{Koide} 
There are constraints from the
universality and it gives the stringent constraints on them. 
The magnitudes of $\epsilon$'s can be ${\mathcal O}(10^{-2})$ \footnote{
In \cite{Koide} the authors use
the very severe constraint for the universality.} 
using the universality which is employed in
Ref.\cite{SmirnovTanimoto}. 
However this constraint can be relaxed
significantly.\cite{SmirnovTanimoto,Ng-Sasaki} 
Thus we can expect rather 
large $\epsilon^{s,m}$ in this kind of models. That is, models in
section \ref{sec:5-2} would give very large $\epsilon^{s,m}$'s.

For the triplet type and the doublet type there are stronger
constraints.
The models in section \ref{sec:5-2} realize 
the doublet type (see Fig.\ref{fig:tree}-(1) with $\tilde{l}_{L}$).
Supersymmetric models with right-handed neutrinos in 
section \ref{sec:5-1} 
give both the triplet type and the singlet type interactions.
In these cases the discussion by the authors in Ref.\cite{Grossman} can be applied
\footnote{Some of the bound indicated in Ref.\cite{Grossman} are
originated from taking into account the interference of the processes 
that the final states are different from each other, but  
such processes can not be added up quantum mechanically. 
We select the bounds given by the processes including the same final
state. To derive the bound for  $\epsilon^{s}_{\mu\tau}$ we use the 
constraint from $\tau\rightarrow \mu ee$ in Ref.\cite{PDG}.
},
that is,
\begin{align}
\epsilon^{s}_{e \tau} \lesssim 3.1 \times 10^{-3}, \quad
\epsilon^{s}_{e \mu} \lesssim 5 \times 10^{-5}, \quad
\epsilon_{\mu \tau}^{s} \lesssim 3.2 \times 10^{-3}.
\end{align}

\section{summary}
In this paper we considered how we can observe the effect of new
physics in a neutrino oscillation experiment.

First we reminded ourselves what we really measure in a neutrino
oscillation experiment. All we really observe are a decay of parent
particle ($\pi$ or $\mu$) at an accelerator cite and an appearance of a
charged lepton at a detector. Neutrinos behave as merely intermediate
states.  Therefore if new physics gives the same amplitudes as those given
by the weak interaction, we cannot distinguish those amplitudes and
hence the transition amplitudes given by new physics can interfere with
those of the weak interaction. It means the effect of new physics can be
amplified quantum-mechanically. That is, we have the effect of not
${\mathcal O}(\lambda^2 )$ but ${\mathcal O}(\lambda)$, where $\lambda$ is an effective
coupling of new physics, in an oscillation experiment.  
We understood also that the all particle states which appear as external lines
including unobserved particles must be the same for the interference
to occur. This statement means that not only the particle species but
also their energy and helicity must be the same.  We have to take into
account both the particle species and a type of an interaction.

Next we derived the transition rate for an appearance of a charged
lepton, which is usually interpreted by the neutrino oscillation.  To
calculate the transition rate in a neutrino factory, we have to consider
 not only particle states in external lines but also a type of new
physics. We separated the effects of new physics in three parts:
(1) We considered new physics affecting the decay process of a parent
particle. If new physics which affects the $\mu$ decay is a $(V-A)(V-A)$
type, eq.\eqref{v-av-a}, then it changes the initial state of neutrinos
from a pure flavor eigenstate to a mixed state given by
eq.\eqref{eq:def-source-state}. On the contrary, for the case of
a $(V-A)(V+A)$ type new interaction, eq.\eqref{v-av+a}, we have a
rather complicated equation for the transition rate, though this effect is
suppressed by $m_e/m_\mu$ due to the difference of the chirality
dependence on the interaction from that on the weak interaction.  
On the other hand, in an oscillation experiment with a conventional beam,
we do not have to worry about the interaction type for the $\pi$ decay,
since 
the energy and the helicity of mouns and neutrinos are fixed due to
the kinematics of the two body decay.
We can parameterize the
effect of new physics using eq.\eqref{eq:def-source-state}.
(2) We studied the effects of new physics in matter.
They are parametrized as eq.\eqref{eq:flavor-changing-matter-effect}
for both oscillation experiments. Their effects also appear linearly
in $\epsilon^m_{\alpha\beta}$ and hence they can give significant
modification on an ``oscillation event''.
(3) We gave a comment how new physics appears at the detector.
The essence is quite similar with that of the decay process 
of a parent particle.
However, to parameterize the effect at the detector we need not only a
magnitude of an elementary process giving a lepton flavor violation but
also a parton distribution function which gives the dependence on the 
neutrino energy. 
We need a precise knowledge about nuclei
so it is very difficult and we ignored these effects.

Then we calculate how many no-oscillation neutrinos are necessary 
to observe the effect of new physics considering only
statistical fluctuation. 
For a neutrino factory we translated this neutrino number into 
the parent muon number.
If we know all the mixing parameters exactly,
then it will be very easy to observe an effect of new
physics, which we parameterized as $\epsilon^{s,m}_{\alpha\beta}$
in eqs.\eqref{eq:flavor-changing-matter-effect} and
\eqref{eq:def-source-state}. 
The optimum setup for the baseline
length and the energy region in a neutrino factory is easily understood
by the high energy expansion of the transition rate as given in
eq.\eqref{eq:HighEnergy_nuenumu_ETau} and $\chi^2$,
eq.\eqref{eq:HighEnergy_nuenumu_ETau_chi2}, for example. 
However, in general, 
we cannot expect that we know the mixing parameters exactly.
If the new physics effect gives a 
similar energy dependence for the transition rate with that of the weak
interaction as eq.\eqref{eq:HighEnergy_nuenumu_ETau}, 
we cannot expect that this effect is observable eventually. 
On the contrary, if the energy dependence is different 
as eq.\eqref{eq:HighEnergy_nuenumu_EMu},
we can expect, even if we have the uncertainties of the mixing parameters,
the new physics effect makes a significant modification to the
transition rate.  
In this case the uncertainties of the mixing parameters
do not make the observability of the new physics effect worse. 
To survey the presence of $\epsilon^{s,m}_{\alpha\beta}$, 
the use of the appearance channel 
of $\nu_\alpha\rightarrow\nu_\beta$ gives the best
sensitivity.  
The sensitivities to new physics effect depend not only on their magnitudes
but also their complex phases. 
More precisely we can observe an effect of
a certain combination of the CP phase $\delta$ and these parameters as given
in for example eq.\eqref{eq:HighEnergy_nuenumu_EMu}. 
The sensitivities are proportional to $|\epsilon|^{2}$.
Naively we can 
expect their significant contribution as long as these combinations
are larger than $\mathcal O(10^{-4})$.

Finally, we gave a discussion about possible new physics and their
consequences to the parameters $\epsilon^{s,m}_{\alpha\beta}$.  
Many models predict large lepton flavor violations to explain neutrino
Majorana masses. 
These violations also affect decays of particles. 
For example, in many models there will be a flavor changing $\mu$ decay.
Also these effects will appear in the matter effects and possibly at the 
detection processes.  Their effects must be taken into account in the
analysis of oscillation experiments if we take those models
seriously. 
Among these effects $\mu\leftrightarrow\tau$ changing effect
is expected to be large to explain the large mixing in the
 atmospheric neutrino anomaly. 
To observe these effect the  disappearance channel of 
$\nu_\mu\rightarrow \nu_{\tau}$ channel are very effective.
This $\nu_\mu\leftrightarrow \nu_{\tau}$ changing effect
may play an important role in $\nu_{\tau}$ appearance experiments
like OPERA.\cite{OPERA}

Incidentally we briefly mention the result of the
 LSND  experiment.\cite{LSND}
In general, its result is interpreted by the neutrino oscillation. 
If its result is partially confirmed by MiniBooNE \cite{MiniBooNE}, 
namely, MiniBooNE see a flavor
changing signal with a small rate which is at the lower end of the LSND
result, then we may interpret the result by the flavor changing decay of
$\pi$ and/or $\mu$. If we take this interpretation, then the expected
signal in a future neutrino-oscillation experiment is  quite different
from that by a four-generation model. We need to keep this possibility in
mind.

\begin{acknowledgments}
 One of the authors (J.S.) would like to thank Y. Kuno, Y. Okada and
 Y. Nir for discussions. 
Two of us (J.S. and N.Y.) also thank K. Inoue for useful comments.   
The work of J.S. is supported in part by the 
Grants-in-Aid for the Ministry of Education, Science, Sports and
 Culture, Government of Japan (No.12047221 and No.12740157).
\end{acknowledgments}

\appendix
\section{Analytic expressions for oscillation probabilities}
\subsection{$P_{\nu^{s}_{e} \rightarrow \nu_{\mu}}$ 
up to $\mathcal{O}$($\delta m^{2}_{21}$, $\epsilon $)}
\begin{align}
P_{\nu_{e}^{s} \rightarrow \nu_{\mu}}
&= s_{23}^{2} s_{2 \times \tilde{13}}^{2}
   \sin^{2} \frac{\lambda_{+}-\lambda_{-}}{4E_{\nu}} L \nonumber
   \\
&  +
  \frac{1}{2} c_{\delta} s_{2 \times 23} s_{2 \times 12} s_{2 \times \tilde
   {13}} \nonumber \\
&  \qquad \times \Biggl[
    \left( c_{\tilde{13}} c_{13 - \tilde{13}}
    \frac{\delta m_{21}^{2}}{\delta m_{21}^{2} c_{12}^
     {2}-\lambda_{-}}
      +
      s_{\tilde{13}} s_{13 - \tilde{13}}
    \frac{\delta m_{21}^{2}}{\lambda_{+}-\delta m_{21}^{2}
    c_{12}^{2}}
    \right)
    \sin^{2} \frac{\delta m_{21}^{2} c_{12}^
     {2}-\lambda_{-}}{4E_{\nu}} L \nonumber \\
&  \qquad \quad -
  \left( c_{\tilde{13}} c_{13 - \tilde{13}}
    \frac{\delta m_{21}^{2}}{\delta m_{21}^{2} c_{12}^
     {2}-\lambda_{-}}
      +
      s_{\tilde{13}} s_{13 - \tilde{13}}
    \frac{\delta m_{21}^{2}}{\lambda_{+}-\delta m_{21}^{2}
    c_{12}^{2}}
    \right)
    \sin^{2} \frac{\lambda_{+}-\delta m_{21}^{2} c_{12}^
     {2}}{4E_{\nu}} L \nonumber \\
&  \qquad \quad +
  \left( c_{\tilde{13}} c_{13 - \tilde{13}}
    \frac{\delta m_{21}^{2}}{\delta m_{21}^{2} c_{12}^
     {2}-\lambda_{-}}
      -
      s_{\tilde{13}} s_{13 - \tilde{13}}
    \frac{\delta m_{21}^{2}}{\lambda_{+}-\delta m_{21}^{2}
    c_{12}^{2}}
    \right)
    \sin^{2} \frac{\lambda_{+}-\lambda_{-}}{4E_{\nu}} L
    \Biggr] \nonumber \\
&  +\frac{1}{4}
  s_{\delta} s_{2 \times 23} s_{2 \times 12} s_{2 \times \tilde{13}}
  \left( c_{\tilde{13}} c_{13 - \tilde{13}}
    \frac{\delta m_{21}^{2}}{\delta m_{21}^{2} c_{12}^
     {2}-\lambda_{-}}
      +
      s_{\tilde{13}} s_{13 - \tilde{13}}
    \frac{\delta m_{21}^{2}}{\lambda_{+}-\delta m_{21}^{2}
    c_{12}^{2}}
    \right) \nonumber \\
&  \qquad \times
  \left(
   \sin \frac{\lambda_{+}-\delta m_{21}^{2} c_{12}^
     {2}}{2E_{\nu}} L
   + \sin \frac{\delta m_{21}^{2} c_{12}^
     {2}-\lambda_{-}}{2E_{\nu}} L
   - \sin \frac{\lambda_{+}-\lambda_{-}}{2E_{\nu}} L
  \right) \\
& \quad
 +
2
s_{23} s_{2 \times \tilde{13}} 
 \left( 
   c_{\delta} {\rm Re}[\epsilon^{s}_{e \mu}]
  + s_{\delta} {\rm Im}[\epsilon^{s}_{e \mu}]
 \right) \nonumber \\
& \qquad \qquad \qquad \qquad
\times
 \left(
  c_{23}^{2} 
  \sin^{2} \frac{\delta m_{21}^{2} c_{12}^{2} - \lambda_{-}}{4E_{\nu}} L
  -
  c_{23}^{2}
  \sin^{2} \frac{\lambda_{+} - \delta m_{21}^{2} c_{12}^{2}}{4E_{\nu}} L
  +
  s_{23}^{2} c_{2 \times \tilde{13}}
  \sin^{2} \frac{\lambda_{+} - \lambda_{-}}{4E_{\nu}} L 
 \right) \nonumber \\
& \quad
+
s_{23} s_{2 \times \tilde{13}} 
 \left( 
   s_{\delta} {\rm Re}[\epsilon^{s}_{e \mu}]
  - c_{\delta} {\rm Im}[\epsilon^{s}_{e \mu}]
 \right) 
 \left(
  c_{23}^{2} 
  \sin \frac{\delta m_{21}^{2} c_{12}^{2} - \lambda_{-}}{2E_{\nu}} L
  +
  c_{23}^{2}
  \sin \frac{\lambda_{+} - \delta m_{21}^{2} c_{12}^{2}}{2E_{\nu}} L
  +
  s_{23}^{2}
  \sin \frac{\lambda_{+} - \lambda_{-}}{2E_{\nu}} L 
 \right) \\
& \quad
 -
s_{23} s_{2 \times 23} s_{2 \times \tilde{13}} 
 \left( 
   c_{\delta} {\rm Re}[\epsilon^{s}_{e \tau}]
  + s_{\delta} {\rm Im}[\epsilon^{s}_{e \tau}]
 \right)
\left(
  \sin^{2} \frac{\delta m_{21}^{2} c_{12}^{2} - \lambda_{-}}{4E_{\nu}} L
  -
  \sin^{2} \frac{\lambda_{+} - \delta m_{21}^{2} c_{12}^{2}}{4E_{\nu}} L
  -
  c_{2 \times \tilde{13}}
  \sin^{2} \frac{\lambda_{+} - \lambda_{-}}{4E_{\nu}} L 
 \right) \nonumber \\
& \quad
 -
\frac{1}{2}
s_{23} s_{2 \times 23} s_{2 \times \tilde{13}} 
 \left( 
   s_{\delta} {\rm Re}[\epsilon^{s}_{e \tau}]
  - c_{\delta} {\rm Im}[\epsilon^{s}_{e \tau}]
 \right)
\left(
  \sin \frac{\delta m_{21}^{2} c_{12}^{2} - \lambda_{-}}{2E_{\nu}} L
  +
  \sin \frac{\lambda_{+} - \delta m_{21}^{2} c_{12}^{2}}{2E_{\nu}} L
  -
  \sin \frac{\lambda_{+} - \lambda_{-}}{2E_{\nu}} L 
 \right)\\
& \quad
 +
 2 s_{23}^{3} s_{2 \times \tilde{13}}^{3}
  \left(  
   \frac{c_{\delta} {\rm Re} [a_{e \mu} ] - s_{\delta}{\rm Im} [a_{e \mu} ]}{4E_{\nu}}L  
  \right)
 \sin \frac{\lambda_{+}- \lambda_{-}}{2E_{\nu}}L \nonumber \\
& \quad 
 +
 s_{2 \times 23} c_{23} s_{2 \times \tilde{13}}
  \Biggl[
     \left( 
       c_{\tilde{13}}^{2}
        \frac{  c_{\delta} {\rm Re} [a_{e \mu} ] 
              - s_{\delta} {\rm Im} [a_{e \mu} ]  }
             {\delta m_{21}^{2} c_{12}^{2} - \lambda_{-} }
        -
       s_{\tilde{13}}^{2}
        \frac{  c_{\delta} {\rm Re} [a_{e \mu} ] 
              - s_{\delta} {\rm Im} [a_{e \mu} ] }
            {\lambda_{+} - \delta m_{21}^{2} c_{12}^{2}}
     \right)
    \sin^{2} \frac{\delta m_{21}^{2} c_{12}^{2} - \lambda_{-}}{4E_{\nu}} L
    \nonumber \\
&  \qquad \qquad 
  -
   \left( 
       c_{\tilde{13}}^{2}
        \frac{  c_{\delta} {\rm Re} [a_{e \mu} ] 
              - s_{\delta} {\rm Im} [a_{e \mu} ]  }
             {\delta m_{21}^{2} c_{12}^{2} - \lambda_{-} }
        -
       s_{\tilde{13}}^{2}
        \frac{ c_{\delta} {\rm Re} [a_{e \mu} ] 
              - s_{\delta} {\rm Im} [a_{e \mu} ] }
            {\lambda_{+} - \delta m_{21}^{2} c_{12}^{2}}
     \right)
   \sin^{2} \frac{\lambda_{+} - \delta m_{21}^{2} c_{12}^{2}}{4E_{\nu}} L 
   \nonumber \\
&  \qquad \qquad 
  + 
    \left( 
       c_{\tilde{13}}^{2}
        \frac{  c_{\delta} {\rm Re} [a_{e \mu} ] 
              - s_{\delta} {\rm Im} [a_{e \mu} ]  }
             {\delta m_{21}^{2} c_{12}^{2} - \lambda_{-} }
       +
       s_{\tilde{13}}^{2}
        \frac{ c_{\delta} {\rm Re} [a_{e \mu} ] 
              - s_{\delta} {\rm Im} [a_{e \mu} ] }
            {\lambda_{+} - \delta m_{21}^{2} c_{12}^{2}}
    \right)
   \sin^{2} \frac{\lambda_{+} - \lambda_{-}}{4E_{\nu}} L
  \Biggr]
\nonumber \\
& \quad
 +\frac{1}{2} s_{2 \times 23} c_{23} s_{2 \times \tilde{13}}
     \left( 
       c_{\tilde{13}}^{2}
        \frac{  s_{\delta} {\rm Re} [a_{e \mu} ] 
              + c_{\delta} {\rm Im} [a_{e \mu} ]  }
             {\delta m_{21}^{2} c_{12}^{2} - \lambda_{-} }
       -
       s_{\tilde{13}}^{2}
        \frac{ s_{\delta} {\rm Re} [a_{e \mu} ] 
              + c_{\delta} {\rm Im} [a_{e \mu} ] }
            {\lambda_{+} - \delta m_{21}^{2} c_{12}^{2}}
    \right) \nonumber \\
&  \qquad \qquad \qquad 
  \times
    \left(
    \sin \frac{\delta m_{21}^{2} c_{12}^{2} - \lambda_{-}}{2E_{\nu}} L
    +
    \sin \frac{\lambda_{+} - \delta m_{21}^{2} c_{12}^{2}}{2E_{\nu}} L
    -
    \sin \frac{\lambda_{+} - \lambda_{-}}{2E_{\nu}} L
    \right) \nonumber \\
& \quad
 +
  4 s_{23}^{3} s_{2 \times \tilde{13}} c_{2 \times \tilde{13}}^{2}
  \left( 
  \frac{  c_{\delta} {\rm Re} [a_{e \mu} ] 
         - s_{\delta} {\rm Im} [a_{e \mu} ] }
        {\lambda_{+} - \lambda_{-}}
  \right)
  \sin^{2} \frac{\lambda_{+} - \lambda_{-}}{4E_{\nu}} L \\
& \quad 
 +
 s_{23} s_{2 \times 23} s_{2 \times \tilde{13}}^{3}
 \left(
  \frac{c_{\delta} {\rm Re} [a_{e \tau} ] - s_{\delta}{\rm Im} [a_{e \tau}] }{4E_{\nu}}L    
 \right)
 \sin \frac{\lambda_{+}- \lambda_{-}}{2E_{\nu}}L \nonumber \\
& \quad 
 -
 s_{23} s_{2 \times 23} s_{2 \times \tilde{13}}
  \Biggl[
   \left( 
       c_{\tilde{13}}^{2}
        \frac{  c_{\delta} {\rm Re} [a_{e \tau} ] 
              - s_{\delta} {\rm Im} [a_{e \tau} ]  }
             {\delta m_{21}^{2} c_{12}^{2} - \lambda_{-} }
        -
       s_{\tilde{13}}^{2}
        \frac{  c_{\delta} {\rm Re} [a_{e \tau} ] 
              - s_{\delta} {\rm Im} [a_{e \tau} ] }
            {\lambda_{+} - \delta m_{21}^{2} c_{12}^{2}}
     \right)
   \sin^{2} \frac{\delta m_{21}^{2} c_{12}^{2} - \lambda_{-}}{4E_{\nu}} L
   \nonumber \\
&  \qquad \qquad 
  -
   \left( 
       c_{\tilde{13}}^{2}
        \frac{  c_{\delta} {\rm Re} [a_{e \tau} ] 
              - s_{\delta} {\rm Im} [a_{e \tau} ]  }
             {\delta m_{21}^{2} c_{12}^{2} - \lambda_{-} }
        -
       s_{\tilde{13}}^{2}
        \frac{ c_{\delta} {\rm Re} [a_{e \tau} ] 
              - s_{\delta} {\rm Im} [a_{e \tau} ] }
            {\lambda_{+} - \delta m_{21}^{2} c_{12}^{2}}
     \right)
   \sin^{2} \frac{\lambda_{+} - \delta m_{21}^{2} c_{12}^{2}}{4E_{\nu}} L 
   \nonumber \\
&  \qquad \qquad 
  + 
    \left( 
       c_{\tilde{13}}^{2}
        \frac{  c_{\delta} {\rm Re} [a_{e \tau} ] 
              - s_{\delta} {\rm Im} [a_{e \tau} ]  }
             {\delta m_{21}^{2} c_{12}^{2} - \lambda_{-} }
       +
       s_{\tilde{13}}^{2}
        \frac{ c_{\delta} {\rm Re} [a_{e \tau} ] 
              - s_{\delta} {\rm Im} [a_{e \tau} ] }
            {\lambda_{+} - \delta m_{21}^{2} c_{12}^{2}}
    \right)
   \sin^{2} \frac{\lambda_{+} - \lambda_{-}}{4E_{\nu}} L
 \Biggr]
 \nonumber \\
& \quad
 -\frac{1}{2} s_{23} s_{2 \times 23} s_{2 \times \tilde{13}}
     \left( 
       c_{\tilde{13}}^{2}
        \frac{  s_{\delta} {\rm Re} [a_{e \tau} ] 
              + c_{\delta} {\rm Im} [a_{e \tau} ]  }
             {\delta m_{21}^{2} c_{12}^{2} - \lambda_{-} }
       -
       s_{\tilde{13}}^{2}
        \frac{ s_{\delta} {\rm Re} [a_{e \tau} ] 
              + c_{\delta} {\rm Im} [a_{e \tau} ] }
            {\lambda_{+} - \delta m_{21}^{2} c_{12}^{2}}
    \right) \nonumber \\
&  \qquad \qquad \qquad 
  \times
    \left(
    \sin \frac{\delta m_{21}^{2} c_{12}^{2} - \lambda_{-}}{2E_{\nu}} L
    +
    \sin \frac{\lambda_{+} - \delta m_{21}^{2} c_{12}^{2}}{2E_{\nu}} L
    -
    \sin \frac{\lambda_{+} - \lambda_{-}}{2E_{\nu}} L
    \right) \nonumber \\
& \quad
 +
  2 s_{23} s_{2 \times 23} s_{2 \times \tilde{13}} c_{2 \times \tilde{13}}^{2}
  \left(
   \frac{  c_{\delta} {\rm Re} [a_{e \tau} ] 
         - s_{\delta} {\rm Im} [a_{e \tau} ] }
        {\lambda_{+} - \lambda_{-}}
  \right)
  \sin^{2} \frac{\lambda_{+} - \lambda_{-}}{4E_{\nu}} L 
\\
& \quad 
 +
 s_{23}^{2} s_{2 \times 23} s_{2 \times \tilde{13}}^{2} c_{2 \times \tilde{13}}
 \left(
  \frac{ {\rm Re} [a_{\mu \tau} ] }{4E_{\nu}}L
 \right)
 \sin \frac{\lambda_{+}- \lambda_{-}}{2E_{\nu}}L \nonumber \\
& \quad 
  -\frac{1}{2} s_{2 \times 23} c_{2 \times 23} s_{2 \times \tilde{13}}^{2}
   \Biggl[
     \left( 
       \frac{ {\rm Re} [a_{\mu \tau} ] }
            {\delta m_{21}^{2} c_{12}^{2} - \lambda_{-} }
        +
       \frac{ {\rm Re} [a_{\mu \tau} ] }
            {\lambda_{+} - \delta m_{21}^{2} c_{12}^{2}}
     \right)
    \sin^{2} \frac{\delta m_{21}^{2} c_{12}^{2} - \lambda_{-}}{4E_{\nu}} L
    \nonumber \\
&  \qquad \qquad \quad  
  -
    \left(
     \frac{ {\rm Re} [a_{\mu \tau} ] }
            {\delta m_{21}^{2} c_{12}^{2} - \lambda_{-} }
        +
       \frac{ {\rm Re} [a_{\mu \tau} ] }
            {\lambda_{+} - \delta m_{21}^{2} c_{12}^{2}}
    \right)
    \sin^{2} \frac{\lambda_{+} - \delta m_{21}^{2} c_{12}^{2}}{4E_{\nu}} L
    \nonumber \\
&  \qquad \qquad \quad  
  +
  \left(
     \frac{ {\rm Re} [a_{\mu \tau} ] }
            {\delta m_{21}^{2} c_{12}^{2} - \lambda_{-} }
        -
       \frac{ {\rm Re} [a_{\mu \tau} ] }
            {\lambda_{+} - \delta m_{21}^{2} c_{12}^{2}}
    \right)
   \sin^{2} \frac{\lambda_{+} - \lambda_{-}}{4E_{\nu}} L
   \Biggr] \nonumber \\
& \quad
  +\frac{1}{4} s_{2 \times 23} s_{2 \times \tilde{13}}^{2}
   \left(
     \frac{ {\rm Im} [a_{\mu \tau} ] }
            {\delta m_{21}^{2} c_{12}^{2} - \lambda_{-} }
        +
       \frac{ {\rm Im} [a_{\mu \tau} ] }
            {\lambda_{+} - \delta m_{21}^{2} c_{12}^{2}}
    \right) \nonumber \\
&  \qquad \qquad \qquad 
  \times
    \left(
    \sin \frac{\delta m_{21}^{2} c_{12}^{2} - \lambda_{-}}{2E_{\nu}} L
    +
    \sin \frac{\lambda_{+} - \delta m_{21}^{2} c_{12}^{2}}{2E_{\nu}} L
    -
    \sin \frac{\lambda_{+} - \lambda_{-}}{2E_{\nu}} L
    \right) \nonumber \\
& \quad
  -2
  s_{23}^{2} s_{2 \times 23} s_{2 \times \tilde{13}}^{2} c_{2 \times \tilde{13}}
   \left(
    \frac{ {\rm Re} [a_{\mu \tau} ] }{\lambda_{+} - \lambda_{-}}
   \right)
  \sin^{2} \frac{\lambda_{+}- \lambda_{-}}{4E_{\nu}}L \\
& \quad 
 -
  s_{23}^{2} s_{2 \times \tilde{13}}^{2} c_{2 \times \tilde{13}}
 \left(
  \frac{ {\rm Re} [a_{e e} ] }{4E_{\nu}}L
 \right)
 \sin \frac{\lambda_{+}- \lambda_{-}}{2E_{\nu}}L 
 +
 2
  s_{23}^{2} s_{2 \times \tilde{13}}^{2}
 \left(
  \frac{ {\rm Im} [a_{e e} ] }{4E_{\nu}}L
 \right)
 \sin^{2} \frac{\lambda_{+}- \lambda_{-}}{4E_{\nu}}L \nonumber \\
& \quad
  +
  2
   s_{23}^{2} s_{2 \times \tilde{13}}^{2} c_{2 \times \tilde{13}}
   \left(
    \frac{ {\rm Re} [a_{e e} ] }{\lambda_{+} - \lambda_{-}}
   \right)
  \sin^{2} \frac{\lambda_{+}- \lambda_{-}}{4E_{\nu}}L \\
& \quad 
 +
  s_{23}^{4} s_{2 \times \tilde{13}}^{2} c_{2 \times \tilde{13}}
 \left(
  \frac{ {\rm Re} [a_{\mu \mu} ] }{4E_{\nu}}L
 \right)
 \sin \frac{\lambda_{+}- \lambda_{-}}{2E_{\nu}}L 
 +
 2
  s_{23}^{4} s_{2 \times \tilde{13}}^{2}
 \left(
  \frac{ {\rm Im} [a_{\mu \mu} ] }{4E_{\nu}}L
 \right)
 \sin^{2} \frac{\lambda_{+}- \lambda_{-}}{4E_{\nu}}L \nonumber \\
& \quad 
  -\frac{1}{4} s_{2 \times 23}^{2} s_{2 \times \tilde{13}}^{2}
   \Biggl[
     \left( 
       \frac{ {\rm Re} [a_{\mu \mu} ] }
            {\delta m_{21}^{2} c_{12}^{2} - \lambda_{-} }
        +
       \frac{ {\rm Re} [a_{\mu \mu} ] }
            {\lambda_{+} - \delta m_{21}^{2} c_{12}^{2}}
     \right)
    \sin^{2} \frac{\delta m_{21}^{2} c_{12}^{2} - \lambda_{-}}{4E_{\nu}} L
    \nonumber \\
&  \qquad \qquad 
  -
    \left(
     \frac{ {\rm Re} [a_{\mu \mu} ] }
            {\delta m_{21}^{2} c_{12}^{2} - \lambda_{-} }
        +
       \frac{ {\rm Re} [a_{\mu \mu} ] }
            {\lambda_{+} - \delta m_{21}^{2} c_{12}^{2}}
    \right)
    \sin^{2} \frac{\lambda_{+} - \delta m_{21}^{2} c_{12}^{2}}{4E_{\nu}} L
    \nonumber \\
&  \qquad \qquad  
  +
  \left(
     \frac{ {\rm Re} [a_{\mu \mu} ] }
            {\delta m_{21}^{2} c_{12}^{2} - \lambda_{-} }
        -
       \frac{ {\rm Re} [a_{\mu \mu} ] }
            {\lambda_{+} - \delta m_{21}^{2} c_{12}^{2}}
    \right)
   \sin^{2} \frac{\lambda_{+} - \lambda_{-}}{4E_{\nu}} L
   \Biggr] \nonumber \\
& \quad
  +\frac{1}{8} s_{2 \times 23}^{2} s_{2 \times \tilde{13}}^{2}
   \left(
     \frac{ {\rm Im} [a_{\mu \mu} ] }
            {\delta m_{21}^{2} c_{12}^{2} - \lambda_{-} }
        +
       \frac{ {\rm Im} [a_{\tau \tau} ] }
            {\lambda_{+} - \delta m_{21}^{2} c_{12}^{2}}
    \right) \nonumber \\
&  \qquad \qquad 
  \times
    \left(
    \sin \frac{\delta m_{21}^{2} c_{12}^{2} - \lambda_{-}}{2E_{\nu}} L
    +
    \sin \frac{\lambda_{+} - \delta m_{21}^{2} c_{12}^{2}}{2E_{\nu}} L
    -
    \sin \frac{\lambda_{+} - \lambda_{-}}{2E_{\nu}} L
    \right) \nonumber \\
& \quad
  -
  2
   s_{23}^{4} s_{2 \times \tilde{13}}^{2} c_{2 \times \tilde{13}}
   \left(
    \frac{ {\rm Re} [a_{\mu \mu} ] }{\lambda_{+} - \lambda_{-}}
   \right)
  \sin^{2} \frac{\lambda_{+}- \lambda_{-}}{4E_{\nu}}L \\
& \quad 
 +
 \frac{1}{4}
  s_{2 \times 23}^{2} s_{2 \times \tilde{13}}^{2} c_{2 \times \tilde{13}}
 \left(
  \frac{ {\rm Re} [a_{\tau \tau} ] }{4E_{\nu}}L
 \right)
 \sin \frac{\lambda_{+}- \lambda_{-}}{2E_{\nu}}L 
 +
 \frac{1}{2}
  s_{2 \times 23}^{2} s_{2 \times \tilde{13}}^{2}
 \left(
  \frac{ {\rm Im} [a_{\tau \tau} ] }{4E_{\nu}}L
 \right)
 \sin^{2} \frac{\lambda_{+}- \lambda_{-}}{4E_{\nu}}L \nonumber \\
& \quad 
  +\frac{1}{4} s_{2 \times 23}^{2} s_{2 \times \tilde{13}}^{2}
   \Biggl[
     \left( 
       \frac{ {\rm Re} [a_{\tau \tau} ] }
            {\delta m_{21}^{2} c_{12}^{2} - \lambda_{-} }
        +
       \frac{ {\rm Re} [a_{\tau \tau} ] }
            {\lambda_{+} - \delta m_{21}^{2} c_{12}^{2}}
     \right)
    \sin^{2} \frac{\delta m_{21}^{2} c_{12}^{2} - \lambda_{-}}{4E_{\nu}} L
    \nonumber \\
&  \qquad \qquad 
  -
    \left(
     \frac{ {\rm Re} [a_{\tau \tau} ] }
            {\delta m_{21}^{2} c_{12}^{2} - \lambda_{-} }
        +
       \frac{ {\rm Re} [a_{\tau \tau} ] }
            {\lambda_{+} - \delta m_{21}^{2} c_{12}^{2}}
    \right)
    \sin^{2} \frac{\lambda_{+} - \delta m_{21}^{2} c_{12}^{2}}{4E_{\nu}} L
    \nonumber \\
&  \qquad \qquad  
  +
  \left(
     \frac{ {\rm Re} [a_{\tau \tau} ] }
            {\delta m_{21}^{2} c_{12}^{2} - \lambda_{-} }
        -
       \frac{ {\rm Re} [a_{\tau \tau} ] }
            {\lambda_{+} - \delta m_{21}^{2} c_{12}^{2}}
    \right)
   \sin^{2} \frac{\lambda_{+} - \lambda_{-}}{4E_{\nu}} L
   \Biggr] \nonumber \\
& \quad
  -\frac{1}{8} s_{2 \times 23}^{2} s_{2 \times \tilde{13}}^{2}
   \left(
     \frac{ {\rm Im} [a_{\tau \tau} ] }
            {\delta m_{21}^{2} c_{12}^{2} - \lambda_{-} }
        +
       \frac{ {\rm Im} [a_{\tau \tau} ] }
            {\lambda_{+} - \delta m_{21}^{2} c_{12}^{2}}
    \right) \nonumber \\
&  \qquad \qquad 
  \times
    \left(
    \sin \frac{\delta m_{21}^{2} c_{12}^{2} - \lambda_{-}}{2E_{\nu}} L
    +
    \sin \frac{\lambda_{+} - \delta m_{21}^{2} c_{12}^{2}}{2E_{\nu}} L
    -
    \sin \frac{\lambda_{+} - \lambda_{-}}{2E_{\nu}} L
    \right) \nonumber \\
& \quad
  -
  \frac{1}{2}
   s_{2 \times 23}^{2} s_{2 \times \tilde{13}}^{2} c_{2 \times \tilde{13}}
   \left(
    \frac{ {\rm Re} [a_{\tau \tau} ] }{\lambda_{+} - \lambda_{-}}
   \right)
  \sin^{2} \frac{\lambda_{+}- \lambda_{-}}{4E_{\nu}}L,
\end{align}
where $\lambda_{\pm}$, $\tilde{\theta}_{13}$ are defined as follows:
\begin{gather}
\lambda_{\pm} \equiv \frac{1}{2} \left[
     \delta m_{31}^{2} + \delta m_{21}^{2} s_{12}^{2} + \bar{a}
      \pm
     \sqrt{ \left\{
            ( \delta m_{31}^{2} - \delta m_{21}^{2} s_{12}^
                {2} ) c_{2 \times 13} - \bar{a}
            \right\}^{2}
            +
            ( \delta m_{31}^{2} - \delta m_{21}^{2} s_{12}^
             {2} )^{2} s_{2 \times 13}^{2}
           } \right], \\
\tan 2 \tilde{\theta}_{13} \equiv 
\frac{s_{2 \times 13} ( \delta m_{31}^{2} - \delta
m_{21}^{2} s_{12}^{2})}{c_{2 \times 13} ( \delta m_{31}^{2} - \delta
m_{21}^{2} s_{12}^{2}) - \bar{a}}. 
\end{gather}
See Ref.\cite{OS} for details of the calculation. 

\subsection{$\Delta P_{\nu^{s}_{\mu} \rightarrow \nu_{\mu}}
 \{\epsilon_{\mu \tau}\}$ 
up to $\mathcal{O}$($\delta m^{2}_{21}$, $\epsilon $)}
\begin{align}
\Delta P_{\nu_{\mu}^{s} \rightarrow \nu_{\mu}} 
  \{ \epsilon_{\mu \tau}^{s,m} \}
&=
- 2 {\rm Re}[\epsilon_{\mu \tau}^{s}]
  s_{2 \times 23}
  \Biggl[
    c_{2 \times 23} s_{\tilde{13}}^{2}
    \left(
     \sin^{2} \frac{\delta m_{21}^{2} c_{12}^{2} - \lambda_{-}}{4E_{\nu}} L 
     -
     \sin^{2} \frac{\lambda_{+} - \delta m_{21}^{2} c_{12}^{2}}{4E_{\nu}} L 
     +
     \sin^{2} \frac{\lambda_{+} - \lambda_{-}}{4E_{\nu}} L
    \right) \nonumber \\
&  \qquad \qquad \qquad \qquad
  + 
    c_{2 \times 23} 
    \sin^{2} \frac{\lambda_{+} - \delta m_{21}^{2} c_{12}^{2}}{4E_{\nu}} L
  -
    s_{\tilde{13}}^{2} (c_{\tilde{13}}^{2} + c_{2 \times 23})
    \sin^{2} \frac{\lambda_{+} - \lambda_{-}}{4E_{\nu}} L
  \Biggr] \nonumber \\
& \quad
 - {\rm Im}[\epsilon_{\mu \tau}^{s}] s_{2 \times 23}
   \left(
    s_{\tilde{13}}^{2} 
    \sin \frac{\delta m_{21}^{2} c_{12}^{2} - \lambda_{-}}{2E_{\nu}} L 
    -
    c_{\tilde{13}}^{2}
    \sin \frac{\lambda_{+} - \delta m_{21}^{2} c_{12}^{2}}{2E_{\nu}} L
   \right) \\
& \quad
+ 
2 \left( \frac{{\rm Re}[a_{\mu \tau}]}{4 E_{\nu}} L \right)
 s_{23}^{2} s_{2 \times 23} 
  \Biggl[
    c_{23}^{2} s_{\tilde{13}}^{2}
      (3 - c_{2 \times \tilde{13}})
       \sin \frac{\delta m_{21}^{2} c_{12}^{2} - \lambda_{-}}{2E_{\nu}} L
       \nonumber \\
&  \qquad \qquad \qquad \qquad \qquad \qquad
    - 
    c_{23}^{2} c_{\tilde{13}}^{2}
      (3 + c_{2 \times \tilde{13}})
       \sin \frac{\lambda_{+} - \delta m_{21}^{2} c_{12}^{2}}{2E_{\nu}} L
       \nonumber \\
&  \qquad \qquad \qquad \qquad \qquad \qquad \qquad \qquad \qquad
   - 2 s_{23}^{2} s_{\tilde{13}}^{2} 
       c_{\tilde{13}}^{2} c_{2 \times \tilde{13}}
       \sin \frac{\lambda_{+} - \lambda_{-}}{2E_{\nu}} L
  \Biggr] \nonumber \\
& \quad 
 + 2 s_{2 \times 23}
   \Biggl[
    \Biggl\{
      2 \frac{{\rm Re}[a_{\mu \tau}]}
             {\delta m_{21}^{2} c_{12}^{2} - \lambda_{-}}
        c_{2 \times 23} s_{\tilde{13}}^{2} 
        (c_{23}^{2} - s_{23}^{2} s_{\tilde{13}}^{2})
     +
     2 \frac{{\rm Re}[a_{\mu \tau}]}
             {\lambda_{+} - \delta m_{21}^{2} c_{12}^{2}}
        c_{2 \times 23} c_{\tilde{13}}^{2} 
        s_{23}^{2} s_{\tilde{13}}^{2} \nonumber \\
&  \qquad \qquad \qquad \qquad \qquad \qquad \qquad \qquad
     +
       \frac{{\rm Re}[a_{\mu \tau}]}{\lambda_{+} - \lambda_{-}}
       s_{23}^{2} s_{2 \times \tilde{13}}^{2} c_{23}^{2}
    \Biggr\}
    \sin^{2} \frac{\delta m_{21}^{2} c_{12}^{2} - \lambda_{-}}{4E_{\nu}} L 
    \nonumber \\
&  \qquad
  -
    \Biggl\{
     2 \frac{{\rm Re}[a_{\mu \tau}]}
             {\delta m_{21}^{2} c_{12}^{2} - \lambda_{-}}
        c_{2 \times 23} s_{\tilde{13}}^{2} 
        s_{23}^{2} c_{\tilde{13}}^{2}
     +
     2 \frac{{\rm Re}[a_{\mu \tau}]}
             {\lambda_{+} - \delta m_{21}^{2} c_{12}^{2}}
        c_{2 \times 23} c_{\tilde{13}}^{2} 
        (c_{23}^{2} - s_{23}^{2} c_{\tilde{13}}^{2}) \nonumber \\
&  \qquad \qquad \qquad \qquad \qquad \qquad \qquad \qquad
   +
       \frac{{\rm Re}[a_{\mu \tau}]}{\lambda_{+} - \lambda_{-}}
       s_{23}^{2} s_{2 \times \tilde{13}}^{2} c_{23}^{2}
    \Biggr\}
    \sin^{2} \frac{\lambda_{+} - \delta m_{21}^{2} c_{12}^{2}}{4E_{\nu}} L 
    \nonumber \\ 
&  \qquad
   +
    \Biggl\{
     2 \frac{{\rm Re}[a_{\mu \tau}]}
             {\delta m_{21}^{2} c_{12}^{2} - \lambda_{-}}
        c_{2 \times 23} s_{\tilde{13}}^{2} 
        s_{23}^{2} c_{\tilde{13}}^{2}
     -
     2 \frac{{\rm Re}[a_{\mu \tau}]}
             {\lambda_{+} - \delta m_{21}^{2} c_{12}^{2}}
        c_{2 \times 23} c_{\tilde{13}}^{2} 
        s_{23}^{2} s_{\tilde{13}}^{2} \nonumber \\
&  \qquad \qquad \qquad \qquad \qquad \qquad \qquad \qquad
   +
       \frac{{\rm Re}[a_{\mu \tau}]}{\lambda_{+} - \lambda_{-}}
       s_{23}^{4} s_{2 \times \tilde{13}}^{2} c_{2 \times \tilde{13}}
    \Biggr\}
    \sin^{2} \frac{\lambda_{+} - \lambda_{-}}{4E_{\nu}} L
   \Biggr].
\end{align}


\end{document}